\documentclass[aip,jap,superscriptaddress,reprint]{revtex4-1}

\usepackage{graphicx}
\usepackage{epstopdf} 
\usepackage{amssymb} 
\usepackage{amsmath} 
\usepackage{dcolumn}

\renewcommand\vec[1]{{\bf #1}}
\newcommand\corr[1]{#1}

\begin{document}

\title{Contact resistances in trigate and FinFET devices in a Non-Equilibrium Green's Functions approach}

\author{L\'eo Bourdet}
\author{Jing Li}
\affiliation{CEA, INAC-MEM, L\_Sim, Grenoble, France}
\affiliation{Univ. Grenoble Alpes, Grenoble, France}

\author{Johan Pelloux-Prayer}
\author{Fran\c cois Triozon}
\author{Mika\"el Cass\'e}
\author{Sylvain Barraud}
\author{S\'ebastien Martinie}
\affiliation{CEA, LETI-MINATEC, Grenoble, France}
\affiliation{Univ. Grenoble Alpes, Grenoble, France}

\author{Denis Rideau}
\affiliation{STMicroelectronics, Crolles, France}

\author{Yann-Michel Niquet}
\email{yniquet@cea.fr}
\affiliation{CEA, INAC-MEM, L\_Sim, Grenoble, France}
\affiliation{Univ. Grenoble Alpes, Grenoble, France}

\begin{abstract}
We compute the contact resistances $R_{\rm c}$ in trigate and FinFET devices with widths and heights in the 4 to 24 nm range using a Non-Equilibrium Green's Functions approach. Electron-phonon, surface roughness and Coulomb scattering are taken into account. We show that $R_{\rm c}$ represents a significant part of the total resistance of devices with sub-30 nm gate lengths. The analysis of the quasi-Fermi level profile reveals that the spacers between the heavily doped source/drain and the gate are major contributors to the contact resistance. The conductance is indeed limited by the poor electrostatic control over the carrier density under the spacers. We then disentangle the ballistic and diffusive components of $R_{\rm c}$, and analyze the impact of different design parameters (cross section and doping profile in the contacts) on the electrical performances of the devices. The contact resistance and variability rapidly increase when the cross sectional area of the channel goes below $\simeq 50$ nm$^2$. We also highlight the role of the charges trapped at the interface between silicon and the spacer material.
\end{abstract}

\maketitle

\section{Introduction}

The electrical performances of field-effect transistors are increasingly limited by the contact resistances $R_{\rm c}$ as the gate length $L$ decreases.\cite{ITRS,Kuhn12} At low drain bias $V_{\rm ds}$, the ``apparent'' contact resistance can be defined as the linear extrapolation to zero gate length of the total resistance $R(L)$ of the device. It embeds, therefore, all contributions that are independent of $L$, namely: {\it i}) the resistance of the metal-semiconductor contact, {\it ii}) the resistance of the highly doped source/drain and of the lowly doped spacers between the contacts and the gate, and {\it iii}) the so-called ``ballistic'' resistance of the device (residual resistance in the absence of scattering mechanisms). Although the latter can be intrinsic to the channel, it is usually \corr{considered part of the contact resistance} as it is independent on the gate length.

There has been a lot of theoretical and experimental works aimed at understanding and improving the transport through the channel of the transistor. At low field, the resistance of the channel can be characterized by the carrier mobility $\mu$, which has been extensively measured and calculated in a variety of silicon structures (bulk,\cite{Jacoboni83} films\cite{Gamiz01,Esseni03a,Uchida03, Casse06,Jin09,Toniutti12,Niquet14,Nguyen14,Niquet15} and wires\cite{Kotlyar04,Jin07,Ramayya08,Poli09a,Poli09c,Persson10,Neophytou11,Lee11,Kim11,Aldegunde11,Luisier11,Akarvardar12,Nguyen13,Coquand13,Coquand13b}). There is much less literature on the contact resistances,\cite{Kim02a,Kim02b,Dixit05,Magnone2008,Tekleab09,Parada11,Park13,Sohn13,An14,Pereira15,Yoon15,Berrada15} even though they have become a major bottleneck for device performances. Most models for the contact resistances are based on drift-diffusion equations needing carrier mobilities as input, which might not be well characterized in the very inhomogeneous source/drain extensions.

In this work, we compute the contact resistances $R_{\rm c}$ of $n$-type, $[110]$ oriented nanowire transistors\cite{Colinge04} with widths $W$ and heights $H$ in the 4 to 24 nm range. We use a Non-Equilibrium Green's Functions (NEGF) approach, which explicitly accounts for confinement and scattering by phonons, surface roughness, and dopants in the inhomogeneous source and drain.\cite{Anantram08} We consider square ($W=H$) and rectangular ($W>H$) trigate devices as well as FinFET ($H>W$) devices with realistic, bulk-like source and drain contacts. We focus on components {\it ii}) and {\it iii}) of the contact resistance (source/drain extensions and ballistic resistances), and show that they indeed represent a significant fraction of the total resistance of these devices. The methodology, based on the $R(L)$ data extrapolation, is a numerical implementation of the transmission line approach\cite{Shokley64,Niquet14} (section \ref{sectionMethodoDevices}). It can be supplemented with a quasi-Fermi level analysis (section \ref{sectionQFL}), which highlights where the \corr{potential drops are located} in the device, and helps to bridge NEGF with Technology Computer Aided Design (TCAD) tools based on drift-diffusion models. The quasi-Fermi level profile shows, in particular, that the contact resistance is dominated by the lowly doped spacers between the source/drain and the channel. We then disentangle the ballistic contribution from the scattering by phonons, surface roughness and impurities (section \ref{sectionBreakdown}). Simple models for the different terms confirm that the contact conductance is most often limited by the (poor) electrostatic control over the carrier density in the spacers. We next discuss the impact of some design parameters (channel cross section and doping profile) on the carrier mobility in the channel, contact resistance and device variability (section \ref{sectionDesign}). We also investigate the effect of charges trapped at the interface between the silicon wire and the spacer material (Si$_3$N$_4$). We finally provide experimental support for the main conclusions of this work (section \ref{sectionExperimental}).

\section{Methodology and devices}
\label{sectionMethodoDevices}

In this section, we discuss the devices, the NEGF approach, the numerical transmission line method for the contact resistance, and a specific example.

\subsection{Devices}

\begin{figure}
\includegraphics[width=0.98\columnwidth]{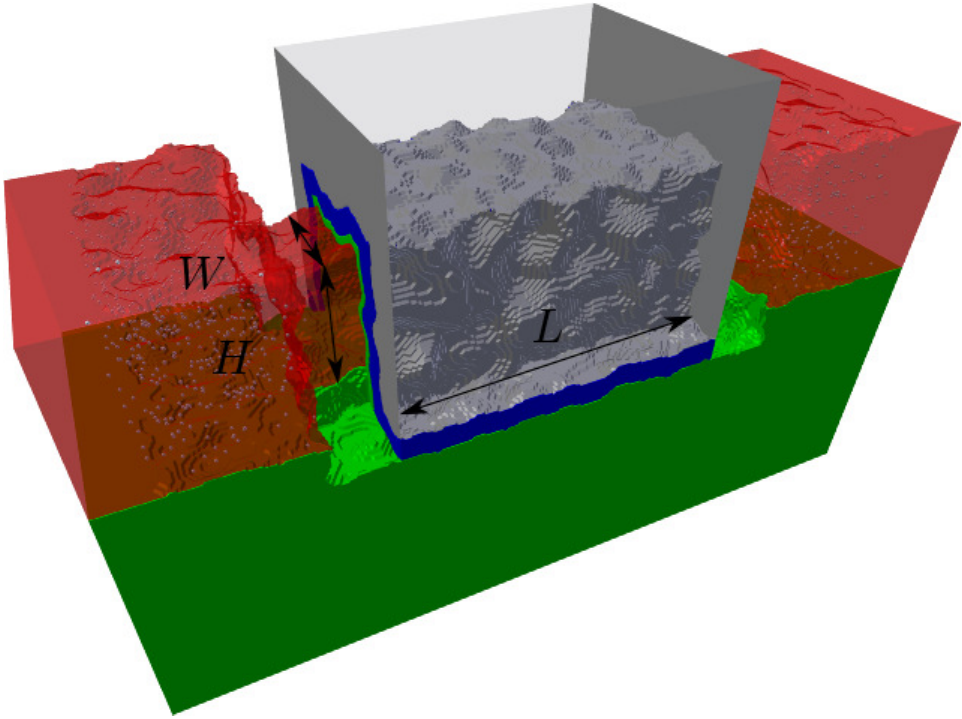}
\caption{A typical $n$-type trigate device with width $W=10$ nm, height $H=10$ nm, and gate length $L=30$ nm. Silicon is in red, SiO$_2$ in green, HfO$_2$ in blue, and the gate in gray (Si$_3$N$_4$ otherwise). The white dots in the source and drain contacts are donor impurities. The silicon substrate is not represented.}
\label{FigDevice}
\end{figure}

A typical device is represented in Fig. \ref{FigDevice}. The channel is a $z=[110]$ oriented rectangular nanowire with width $W$ and height $H$, etched in a $(001)$ Silicon-On-Insulator (SOI) layer.\cite{Coquand13,Coquand13b} It is lying on a 25 nm thick buried oxide (BOX) and a $n$-doped Si substrate (donor concentration $N_d=10^{18}$ cm$^{-3}$). The gate stack covers the top $(001)$ and side $(1\bar{1}0)$ facets of the nanowire. It is made of a 0.8 nm thick layer of SiO$_2$ (dielectric constant $\varepsilon=3.9$) and a 2.2 nm thick layer of HfO$_2$ ($\varepsilon=20$). The source and drain are modeled as films with thickness $t_{\rm SD}\ge H+7$ nm. \corr{Periodic boundary conditions are, therefore, applied along $y=[1\bar{1}0]$, on both potential and wave functions.\cite{NoteGeo} The width of the unit cell is $W_{\rm SD}\ge W+14$ nm. The length of the source and drain films included in the simulation domain is $L_{\rm SD}=16$ nm.} The source and drain are separated from the gate by Si$_3$N$_4$ spacers with length $L_{\rm sp}=6$ nm ($\varepsilon=7.5$).

Surface roughness (SR) and charge disorders are explicitly included in the geometries. For SR, we use a Gaussian auto-correlation function model\cite{Goodnick85} with different correlation lengths $\Lambda$ and rms $\Delta$ on different facets: $\Lambda=1.5$ nm and $\Delta=0.25$ nm on the bottom interface with the BOX, $\Lambda=1.5$ nm and $\Delta=0.35$ nm on the top $(001)$ facet, and $\Lambda=2.0$ nm, $\Delta=0.45$ nm on the side $(1\bar{1}0)$ facets. We also include Remote Coulomb Scattering (RCS) in the gate stack as a distribution of positive charges at the SiO$_2$/HfO$_2$ interface with density $n_{\rm RCS}=2\times 10^{13}$ cm$^{-2}$. The SR parameters for the top and bottom interfaces are chosen to reproduce the experimental mobility in planar FDSOI devices (along the lines of Ref. \onlinecite{Nguyen14}). The SR parameters of the side facets as well as $n_{\rm RCS}$ are chosen to reproduce the experimental mobility in $W=10\times H=10$ nm trigate devices (see section \ref{sectionQFL}). The sidewalls appear rougher than the top and bottom interfaces because they are etched in the SOI film. 

\begin{figure}
\includegraphics[width=0.98\columnwidth]{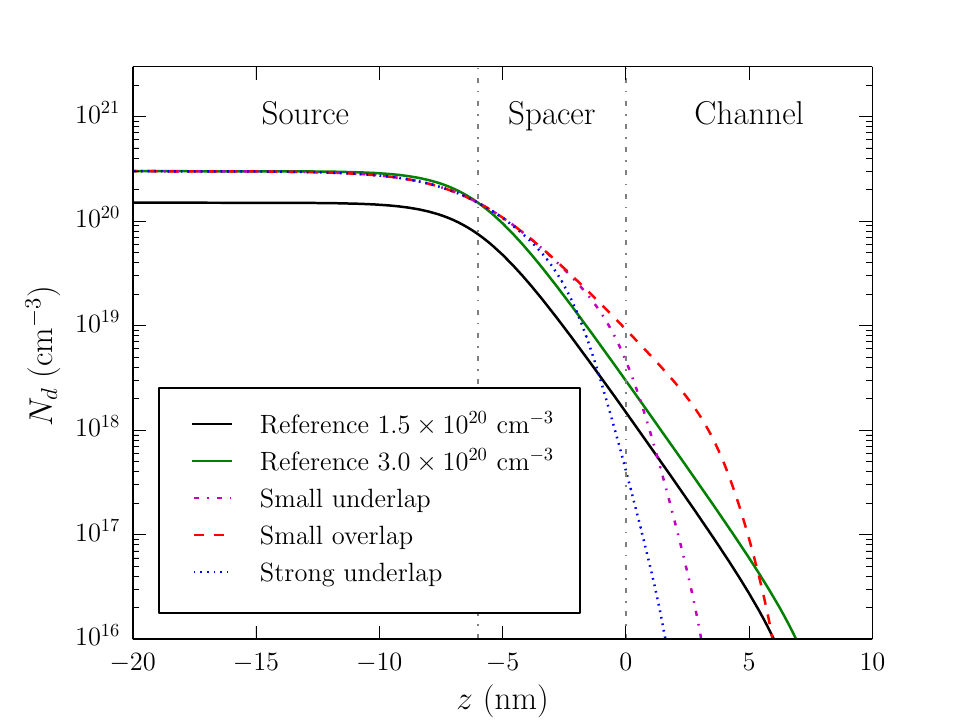}
\caption{Target doping profiles on the source side (the profiles are symmetric on the drain side).}
\label{figDoping}
\end{figure}

The source, drain and spacers are $n$-doped with point charges as models for ionized donor impurities. These point charges are randomly scattered using the doping profiles shown in Fig. \ref{figDoping} as target distribution functions. The ``Reference'' profiles have $N_d=1.5\times 10^{20}$ cm$^{-3}$ or $N_d=3\times 10^{20}$ cm$^{-3}$ in the contacts, and a single decay length $\lambda_1=3$ nm/decade under the spacers and channel. The other profiles will be discussed in detail in section \ref{sectionDoping}. Near the edges of the simulation box the point charge distributions are mixed with the target continuous background charge distributions in order to smooth the variations of the potential at the boundaries where the contact self-energies\cite{Anantram08} are plugged in.\cite{noteDoping}

\subsection{The NEGF approach}

The current is computed in a self-consistent NEGF framework,\cite{Anantram08,Stefanucci} within the effective mass approximation (EMA).\cite{Bastard} The NEGF equations are solved on a finite differences grid (with 2\,\AA\ step), in a fully coupled mode space approach (160 to 352 modes depending on the cross section). Electron-phonon scattering is described by an acoustic deformation potential $D_{\rm ac}=14.6$ eV,\cite{Esseni03a} and by the three $f$ and three $g$ inter-valley processes of Ref. \onlinecite{Jacoboni83}. \corr{Technical details about the electron-phonon self-energies and solution of the NEGF equations can be found in the Appendix of Ref. \onlinecite{Niquet14}.} The $n$-doped substrate is biased at back-gate voltage $V_{\rm bg}=0$ V and is treated semi-classically.\cite{Sze}

\corr{Electron-electron interactions are treated in a self-consistent ``Schr\"odinger-Poisson'' approximation, i.e. the carriers are moving in the potential created by the average density of conduction band electrons in silicon. The effects of valence band electrons in silicon and other materials are accounted for by the dielectric constants introduced in Poisson's equation. This approximation neglects Coulomb correlations} that might, however, become important in nanoscale devices.\cite{Fischetti09} As a step further, we have introduced (in a few test cases) a $GW$-like self-energy accounting for the long-range correlations brought by the dielectric mismatch between the different materials, supplemented with a local density approximation for the short-range exchange-correlation effects.\cite{Jin09,Li09,Niquet15,Lavieville15} This has a moderate impact on the contact resistances ($5-10\%$ increase). Since the calculation of this $GW$-like self-energy is very demanding, these corrections have been neglected in the following. 

One of the main advantages of NEGF is that all structural scattering mechanisms (SR, impurity and RCS scattering) are treated explicitly. There is no need for models for the interactions with these disorders. As a matter of fact, such models are still missing under the spacers, which are very inhomogeneous but critical regions on the resistive path (see sections \ref{sectionQFL} and \ref{sectionBreakdown}). Also, at variance with most previous NEGF calculations,\cite{Poli09a,Poli09c,Luisier11,Aldegunde11,Nguyen13,Seoane09,Martinez12,Berrada15} the source and drain contacts are wide enough with respect to the channel to act as bulk reservoirs (with a 3D-like density of states). This is essential for a quantitative description of the contact resistances. The convergence \corr{with respect to $t_{\rm SD}$, $W_{\rm SD}$ and $L_{\rm SD}$ has}, in particular, been carefully checked.

\subsection{Methodology}
\label{subsectionMethodology}

At low drain bias $V_{\rm ds}$, the total resistance of the device is expected to increase linearly with gate length $L$:\cite{Datta97}
\begin{equation}
R(L)=\frac{V_{\rm ds}}{I_{\rm ds}}=R_0+R_{\rm s}+R_{\rm d}+\frac{L}{n_{\rm 1d}\mu e}\,,
\label{eqRL}
\end{equation}
where $I_{\rm ds}$ is the drain current, $R_{\rm s}$ is the diffusive resistance of the source, $R_{\rm d}$ is the diffusive resistance of the drain, $\mu$ is the carrier mobility and $n_{\rm 1d}$ the carrier density per unit length in the channel, and $R_0$ is the ballistic resistance of the device (residual resistance in the absence of scattering mechanisms).\cite{Datta97,Shur02} At zero temperature, the ballistic conductance of a homogeneous conductor, $G_0=R_0^{-1}=(12.9{\rm\ k}\Omega/N)^{-1}$, is limited by the number $N$ of 1D sub-bands carrying current. At room temperature, $G_0$ reads, assuming Maxwell-Boltzmann statistics and a single transport mass $m^*$ for all sub-bands:\cite{Datta97,Nguyen13}
\begin{equation}
G_0=\frac{n_{\rm 1d}e^2}{\sqrt{2\pi m^*kT}}\,.
\label{eqR0}
\end{equation}
It is, therefore, proportional to the carrier density in the conductor -- we will come back to that point in section \ref{sectionBreakdown}. 

The ``apparent'' contact resistance $R_{\rm c}=R_0+R_{\rm s}+R_{\rm d}$ can hence be extracted from the linear extrapolation of $R(L)$ down to $L=0$ (an approach known as the transmission line method\cite{Shokley64}). For that purpose, we prepare a series of devices with lengths $L=30$ nm and $L=60$ nm (and up to $L=90$ nm), sharing the same contacts (source/drain/spacers geometry and dopant distributions). In order to limit the noise on the $R(L)$ data that might arise from different disorders at different $L$'s, we repeat the 30 nm long sample of surface roughness and RCS charges along the $60$ and $90$ nm long channels.\cite{Niquet14} We can extract that way the resistance of the chosen contact geometry and dopant distribution very accurately. We then average the contact resistance over 3 to 8 surface roughness profiles and dopant distributions in order to assess local variability. 

In order to ease the comparison between devices with different cross sections, the carrier densities $n_{\rm 2d}=n_{\rm 1d}/W_{\rm eff}$ and the resistances $\bar{R}=RW_{\rm eff}$ are normalized to the total effective width of the channel $W_{\rm eff}=W+2H$. The drain bias is $V_{\rm ds}=10$ mV, and the threshold voltage $V_{\rm t}$ is extracted from the $Y=I_{\rm ds}/\sqrt{g_m}$ function (where $g_m=dI_{\rm ds}/dV_{\rm gs}$ is the transconductance and $V_{\rm gs}$ the gate bias).\cite{Ghibaudo88} There are no significant short-channel effects (in particular, no $V_{\rm t}$ roll-off) \corr{in any of} the investigated devices.

\subsection{Example: The $W=10\times H=10$ nm trigate}
\label{subsectionExampleRL}

In this paragraph, we illustrate the above methodology on a specific example: The $W=10\times H=10$ nm trigate device with the ``Reference $1.5\times 10^{20}$ cm$^{-3}$'' doping profile of Fig. \ref{figDoping}.

\begin{figure}
\includegraphics[width=0.98\columnwidth]{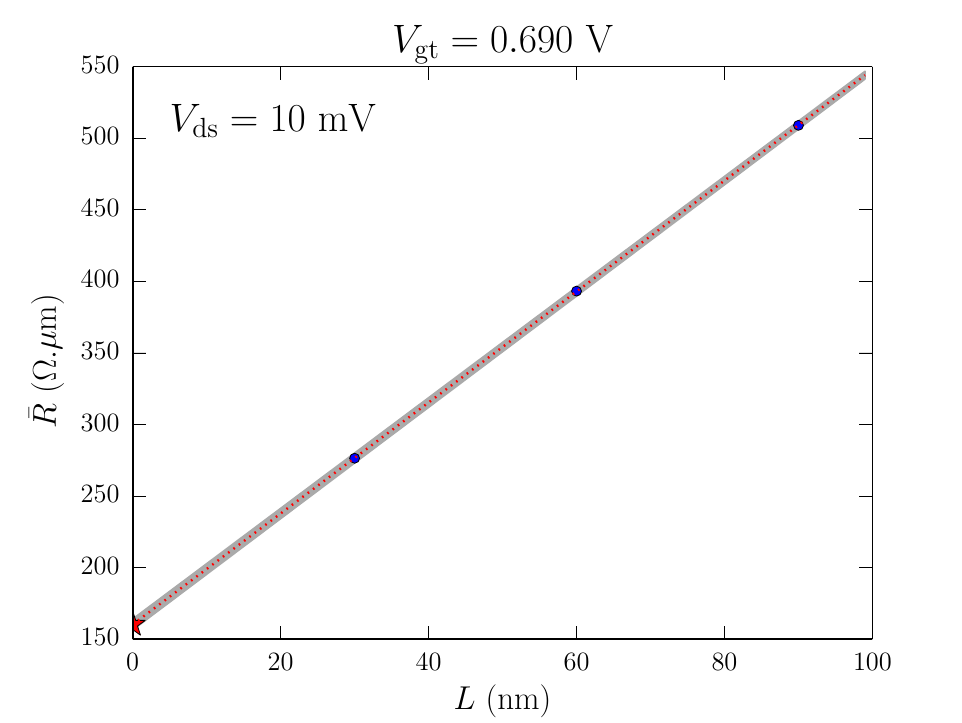}
\caption{Resistance $R(L)$ as a function of gate length $L$ for a $W=10\times H=10$ nm trigate device with the ``Reference $1.5\times 10^{20}$ cm$^{-3}$'' doping profile of Fig. \ref{figDoping}. The gate voltage is $V_{\rm gt}=V_{\rm gs}-V_{\rm t}=0.69$ V, which corresponds to a carrier density $\langle n_{\rm 2d}\rangle_{\rm ch}\simeq 9\times 10^{12}$ cm$^{-2}$ in the channel. The shaded gray area is the 90\% confidence interval for the linear regression (dotted red line), and the red star is the extrapolated $\bar{R}_{\rm c}\equiv\bar{R}(L=0)$.}
\label{figRL}
\end{figure}

\begin{figure}
\includegraphics[width=0.98\columnwidth]{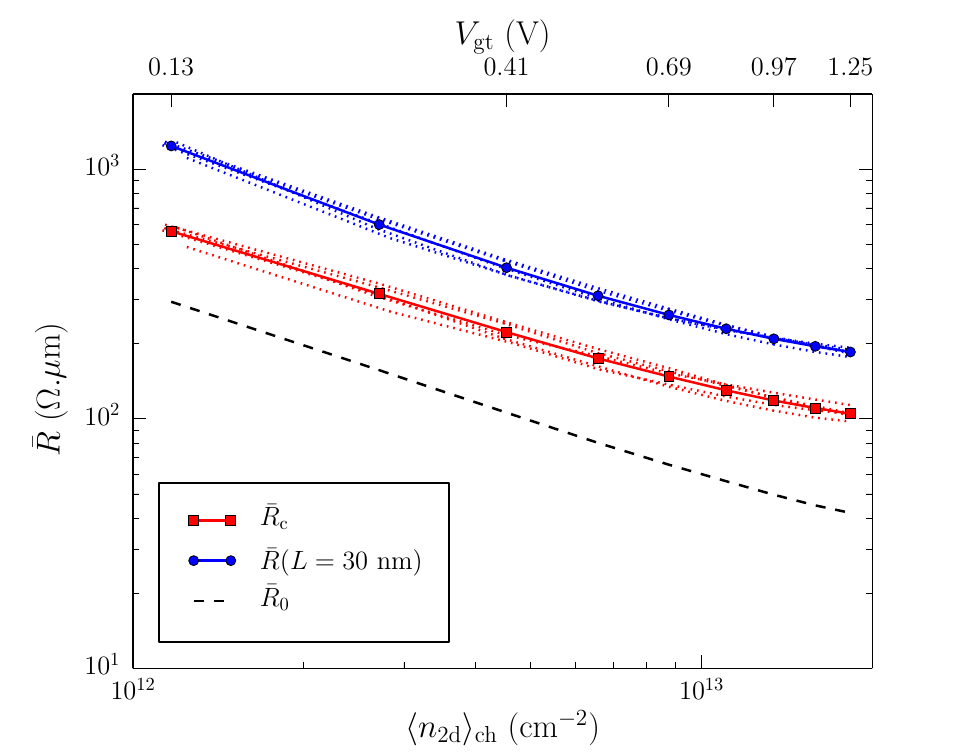}
\caption{Contact resistance $\bar{R}_{\rm c}$ as a function of the gate overdrive (upper axis)/carrier density (lower axis), in a $W=10\times H=10$ nm trigate device with the ``Reference $1.5\times 10^{20}$ cm$^{-3}$'' doping profile of Fig. \ref{figDoping}. Data for five devices (dotted lines, each corresponding to different disorders in the channel and contacts), and their average (solid line) are represented. $\bar{R}_{\rm c}$ is compared to the total resistance $\bar{R}(L=30\text{ nm})$ of the 30 nm long devices, and to the resistance $\bar{R}_0$ of the corresponding ballistic device.}
\label{figR10x10}
\end{figure}

The total resistance of a series of devices is plotted as a function of the channel length in Fig. \ref{figRL}, at gate overdrive $V_{\rm gt}=V_{\rm gs}-V_{\rm t}=0.69$ V. As expected, the resistance scales linearly with length in the inversion regime. The contact resistance $\bar{R}_{\rm c}$ (red star on Fig. \ref{figRL}) and the channel mobility $\mu$ can therefore be extracted from these data. $\bar{R}_{\rm c}$ is plotted as a function of the gate overdrive $V_{\rm gt}$ and carrier density $\langle n_{\rm 2d}\rangle_{\rm ch}$ in the channel in Fig. \ref{figR10x10}. $\langle n_{\rm 2d}\rangle_{\rm ch}$ is defined as the average carrier density in the central, 30 nm long segment of the 60 nm long channel. $\bar{R}_{\rm c}$ follows an approximate $\langle n_{\rm 2d}\rangle_{\rm ch}^{-\beta}$ (or $V_{\rm gt}^{-\beta}$) trend (with $\beta$ close to 1). As the conductivity is proportional to the carrier density in the simplest Drude model, this trend suggests that the contact resistance is dominated by the near channel (spacers) region.\cite{Rideau14} This will be confirmed by a quasi-Fermi level analysis in the next section. 

Data for five devices with different disorders are represented in Fig. \ref{figR10x10}. The local variability due to fluctuations in surface roughness and dopant distributions remains weak in $10\times 10$ nm devices. $\bar{R}_{\rm c}$ is compared to the total resistance $\bar{R}(L=30\text{ nm})$ of the 30 nm long devices, and to the the resistance $\bar{R}_0$ of the corresponding ``ballistic'' device (neither phonons, nor SR and RCS, with a continuous background dopant distribution in the contacts). The contact resistance is around half the total resistance in a 30 nm long device -- it would hence be as large as two-thirds the total resistance in a $\sim 15$ nm long channel. It is, moreover, much larger than the ballistic resistance (especially in the strong inversion regime), which is a component and lower bound of $\bar{R}_{\rm c}$. The optimization of the source/drain/spacers region is, therefore, at least as important as the design of the channel in these transistors.

\section{A quasi-Fermi level analysis of the resistive path}
\label{sectionQFL}

The above conclusions can be further supported by a quasi-Fermi level analysis, which highlights where the potential drops in the system. We first define the quasi-Fermi level in the NEGF framework, then discuss limiting cases (ballistic and diffusive conductors), and finally make a detailed analysis on the trigate device of paragraph \ref{subsectionExampleRL}.

\subsection{Definition}

Drift-diffusion models\cite{Sze} assume that the carriers are in local equilibrium with a quasi-Fermi level $\varepsilon_f(\vec{r})$. The current density in the device is then proportional to the gradient of $\varepsilon_f(\vec{r})$:
\begin{equation}
\vec{j}=n\mu\nabla_\vec{r}\varepsilon_f\,.
\label{eqDD}
\end{equation}
In the NEGF framework, the carriers can be driven far out-of-equilibrium, so that the quasi-Fermi level is, in general, ill-defined. The distribution of electrons can instead be characterized by the local distribution function:
\begin{equation}
f(\vec{r},E)=D_{\rm occ}(\vec{r},E)/D(\vec{r},E)\,,
\end{equation}
where $D_{\rm occ}(\vec{r},E)={\rm Im}\,G^{<}(\vec{r},\vec{r},E)/(2\pi)$ is the density of occupied states (with $G^<$ the lesser Green's function), and $D(\vec{r},E)=-{\rm Im}\,G^r(\vec{r},\vec{r},E)/\pi$ is the total density of states (with $G^r$ the retarded Green's function).\cite{Anantram08} Yet a detailed analysis shows that $f(\vec{r},E)$ remains close to a Fermi function at low bias. In that limit, we can define the quasi-Fermi level $\varepsilon_f(\vec{r})$ as the chemical potential that reproduces the NEGF carrier density $n(\vec{r})$ assuming local Fermi-Dirac statistics:
\begin{equation}
n(\vec{r})=\int dE\, D(\vec{r},E)\, f\left[\left(E-\varepsilon_f(\vec{r})\right)/kT\right]\,,
\label{eqQFL}
\end{equation}
where $f(x)=1/(1+e^x)$ is the reduced Fermi function. 

The quasi-Fermi level analysis is an efficient way to bridge NEGF with the drift-diffusion models widely used in industrial TCAD tools. For practical purposes, we define a one-dimensional quasi-Fermi level $\varepsilon_f(z)$ in the device from the relation:
\begin{equation}
n_{\rm 1d}(z)=\int dE\, D_{\rm 1d}(z, E)\, f\left[\left(E-\varepsilon_f(z)\right)/kT\right]\,,
\label{eqQFLz}
\end{equation}
where $n_{\rm 1d}(z)$ and $D_{\rm 1d}(z, E)$ are the NEGF carrier density and density of states per unit length, which are the integrals of $n(\vec{r})$ and $D(\vec{r},E)$ in the cross section at $z$. 

\subsection{Limiting cases: Ballistic and diffusive conductors}
\label{subsectionCases}

In this paragraph, we address two limiting cases: the ballistic and diffusive conductors. We discuss how the quasi-Fermi level drops in these conductors, in order to provide guidelines for the analysis of trigate and FinFET devices.

\corr{Although the carrier distribution in a ballistic conductor is definitely not a Fermi-Dirac distribution,\cite{Rhew02} the solution of Eq. (\ref{eqQFLz}) remains unambiguously defined. Let us consider a ballistic transistor in the ON state, at low $V_{\rm ds}$ and low temperature.} The potential in such a transistor can be modeled in a first approximation as a square barrier. Carriers with positive group velocities (labeled as wave vectors $k_z>0$ for simplicity) are injected from the source and are in equilibrium with the chemical potential $\mu_{\rm s}$. Carriers with negative group velocities ($k_z<0$) are injected from the drain and are in equilibrium with the chemical potential $\mu_{\rm d}=\mu_{\rm s}-eV_{\rm ds}$. If the carrier density is much lower in the channel than in the source and drain (which is usually the case), most of the electrons are backscattered by the barrier, and only a small fraction flows in the channel. The quasi-Fermi level hence tends to $\mu_{\rm s}$ in the source, and to $\mu_{\rm d}$ in the drain. In the channel, there is a population of carriers with positive group velocities flowing from the source and in equilibrium with $\mu_{\rm s}$, and a population of carriers with negative group velocities flowing from the drain and in equilibrium with $\mu_{\rm d}$. Hence, all $\pm k_z$ states are filled below $\mu_{\rm d}$, while only the $k_z>0$ states are occupied between $\mu_{\rm d}$ and $\mu_{\rm s}$. If the density of states is approximately constant in the  $[\mu_{\rm d}, \mu_{\rm s}]$ range, the quasi-Fermi level in the channel must be $\varepsilon_f=(\mu_{\rm s}+\mu_{\rm d})/2$ in order to match the carrier density with a Fermi-Dirac distribution.

This argument shows that the quasi-Fermi level essentially drops at both ends of the channel in a ballistic device.\cite{Buttiker85,Datta} Ballistic device simulations confirm that this is indeed the case. The width of the drop on each end of the channel depends, in particular, on the shape of the barrier and on the temperature. The drop is typically faster in the strong than in the weak inversion regime, where the quasi-Fermi level can show significant variations over the whole channel in short devices.

In a diffusive conductor on the other hand, the current must be sustained by a finite electric field. The quasi-Fermi level shall show the same variations as the potential if the density is approximately constant, i.e. it shall drop linearly in the channel at low bias.

\corr{As a concluding remark, we would like to emphasize that the above analysis only applies at low bias, and might break down at large $V_{\rm ds}$ in both ballistic\cite{Rhew02} and diffusive cases.}

\subsection{Application to the $W=10\times H=10$ nm trigate}
\label{subsectionExampleQFL}

\begin{figure}
\includegraphics[width=0.98\columnwidth]{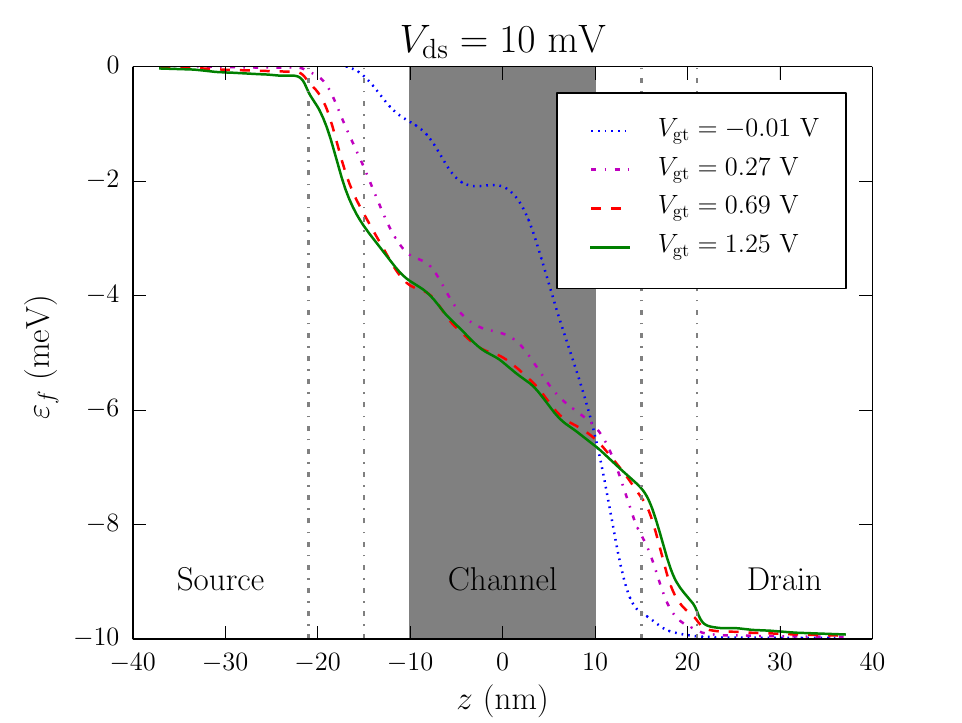}
\caption{Quasi-Fermi level in a 30 nm long, $W=10\times H=10$ nm trigate device with the ``Reference $1.5\times 10^{20}$ cm$^{-3}$'' doping profile of Fig. \ref{figDoping}, at different gate overdrives. The drain bias is $V_{\rm ds}=10$ mV. The shaded gray area delimits the segment where the slope $d\varepsilon_f/dz$ of the quasi-Fermi level is computed.}
\label{figQFL}
\end{figure}

The quasi-Fermi level computed in the same device as in paragraph \ref{subsectionExampleRL} is plotted in Fig. \ref{figQFL} at different gate overdrives (drain bias $V_{\rm ds}=10$ mV).

As expected from the above discussion, the quasi-Fermi level tends to zero in the source (reference chemical potential $\mu_{\rm s}=0$), and to $\mu_{\rm d}=-eV_{\rm ds}$ in the drain. Moreover, the quasi-Fermi level is almost constant in the source and drain, and drops under the spacers and in the channel. These regions are, therefore, the most resistive parts of the device. The resistance of the heavily doped, bulk source and drain is negligible because the density of carriers and cross sectional area are much larger there than in the channel, even though the mobility is expected to be very low in the contacts ($<100$ cm$^2$/V/s).\cite{Jacoboni77}

\begin{figure}
\includegraphics[width=0.98\columnwidth]{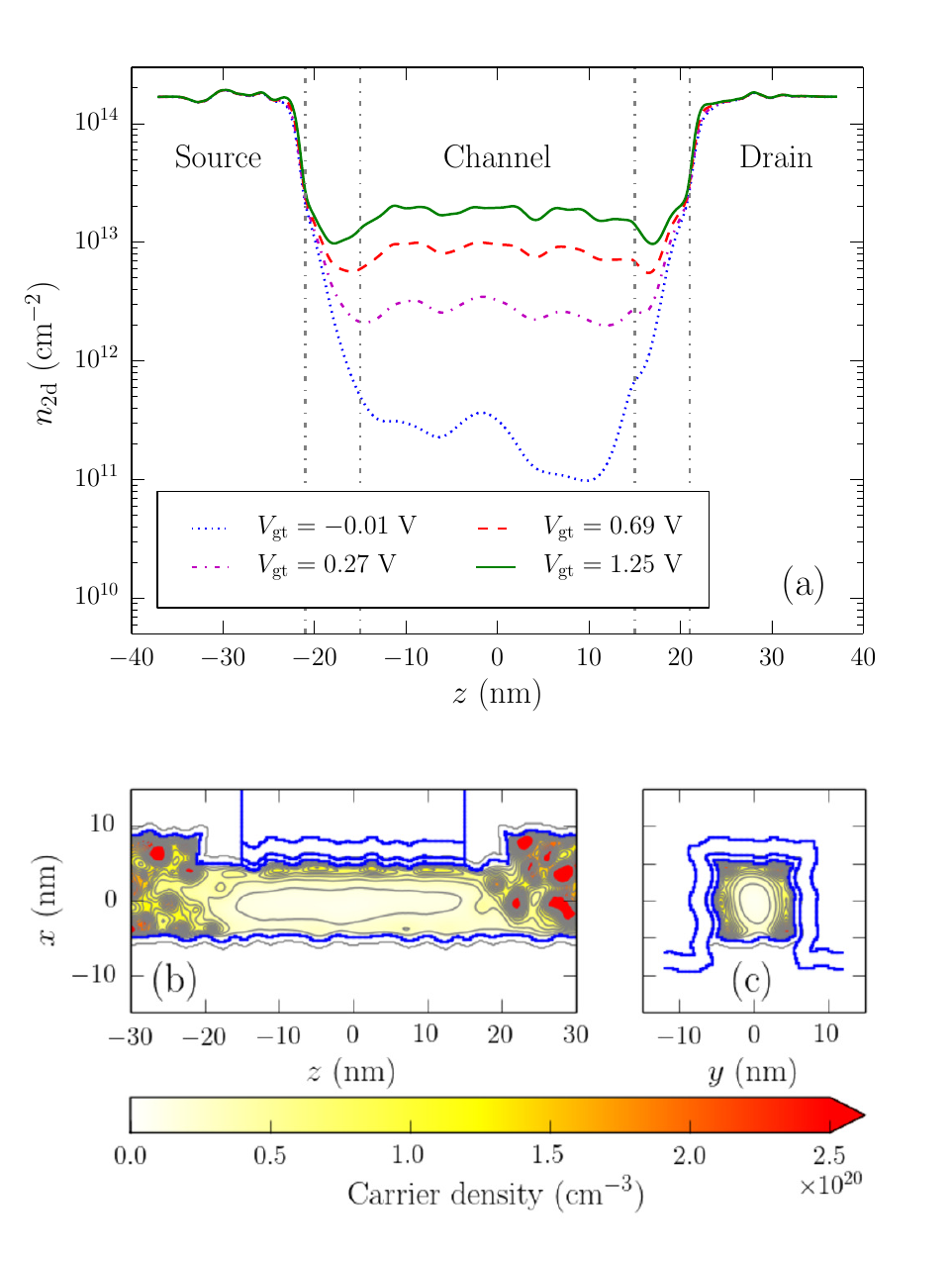}
\caption{(a) Carrier density $n_{\rm 2d}$ in a 30 nm long, $W=10\times H=10$ nm trigate device with the ``Reference $1.5\times 10^{20}$ cm$^{-3}$'' doping profile of Fig. \ref{figDoping}, at different gate overdrives. (b, c) 3D carrier density in a longitudinal and transverse cross section through the center of the channel ($V_{\rm gt}=1.25$ V). The strong disorder in the contacts is due to dopants.}
\label{figDensity}
\end{figure}

The quasi-Fermi level profile is, nonetheless, strongly dependent on the gate bias. Near or under the threshold, the transport is strongly limited by the bell-shaped barrier under the gate. The density is, moreover, very inhomogeneous in the channel (see Fig. \ref{figDensity}). The quasi-Fermi level drops unevenly in the channel, and no clear distinction can be made between a ballistic and a diffusive (scattering-limited) component.

The picture is clearer at moderate to strong inversion. The carrier density is much more homogeneous under the gate. The quasi-Fermi level drops quasi-linearly in the channel, and exhibits two steps under the spacers. We can conjecture from paragraph \ref{subsectionCases} that {\it i}) the transport in the channel is diffusive (the carrier mobility $\mu$ shall hence be proportional to slope of the quasi-Fermi level under the gate) ; {\it ii}) the drop of the quasi-Fermi level under the spacers is proportional to the contact resistance $\bar{R}_{\rm c}$ (including the ballistic component $\bar{R}_0$). 

\begin{figure}
\includegraphics[width=0.98\columnwidth]{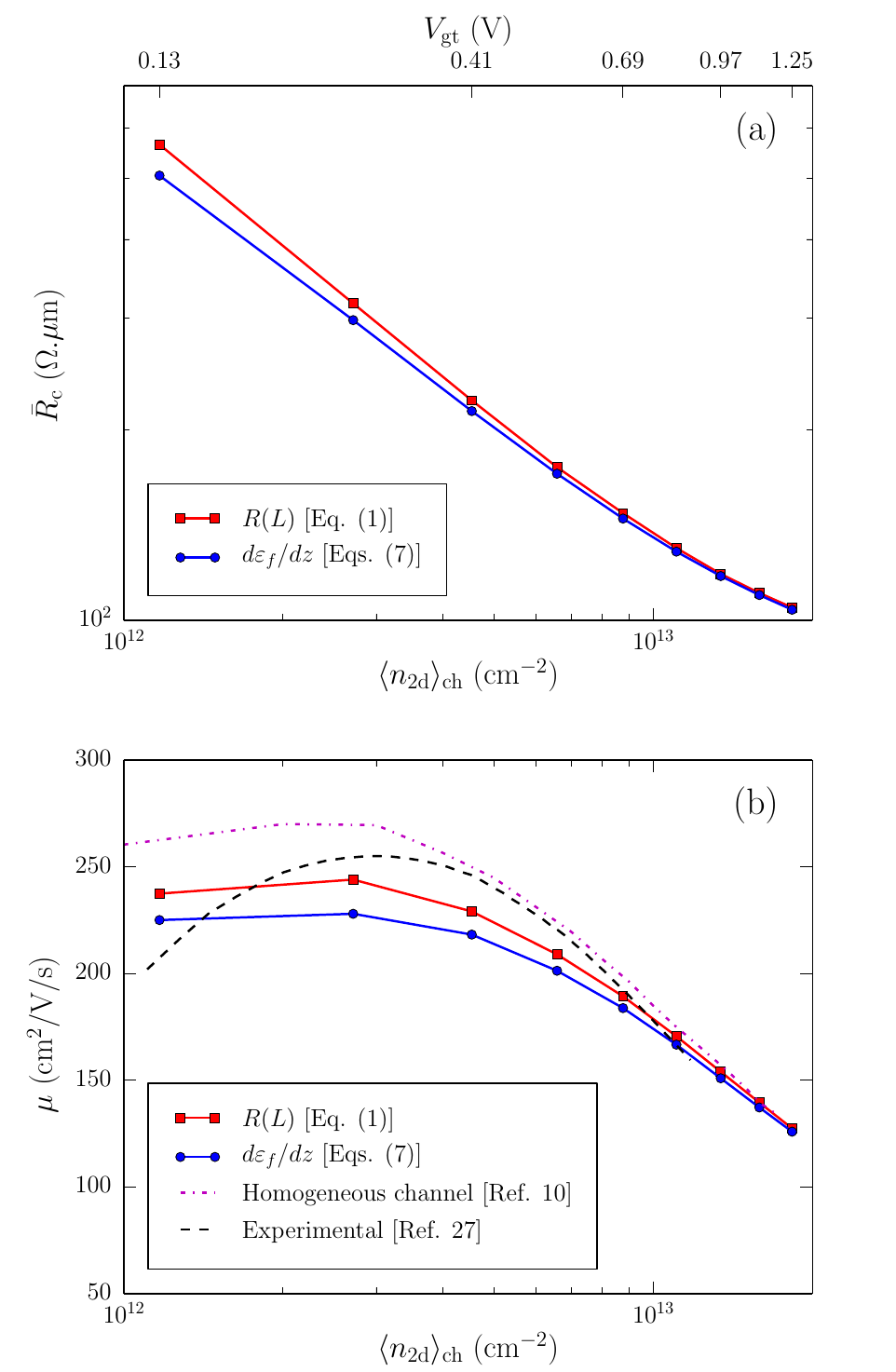}
\caption{(a) Contact resistance $\bar{R}_{\rm c}$ and (b) carrier mobility $\mu$ in the channel of a $W=10\times H=10$ nm trigate device with the ``Reference $1.5\times 10^{20}$ cm$^{-3}$'' doping profile of Fig. \ref{figDoping}, extracted from: {\it i}) a linear regression on the $R(L)$ data [Eq. (\ref{eqRL})], and {\it ii}) the slope $d\varepsilon_f/dz$ of the quasi-Fermi level under the gate of the 30 nm long device [Eqs. (\ref{eqRqfl2})]. The data are averaged over five devices as in Fig. \ref{figR10x10}. The mobility computed in homogeneous nanowire channels without spacers and S/D contacts is also plotted (see Ref. \onlinecite{Niquet14}) along with experimental data (Ref. \onlinecite{Coquand13b}).}
\label{figR10x10qfl}
\end{figure}

To support these conclusions, we have recomputed the channel mobility $\mu$ and contact resistance $\bar{R}_{\rm c}$ from the relations:
\begin{subequations}
\label{eqRqfl2}
\begin{align}
&\frac{1}{\langle n_{\rm 1d}\rangle_{\rm ch}\mu e}=\frac{1}{eI_{\rm ds}}\left\langle\frac{d\varepsilon_f}{dz}\right\rangle_{\rm ch} \\
&\bar{R}_{\rm c}=\frac{V_{\rm ds}}{I_{\rm ds}}-\frac{L}{\langle n_{\rm 1d}\rangle_{\rm ch}\mu e}\,,
\end{align}
\end{subequations}
where $\langle d\varepsilon_f/dz\rangle_{\rm ch}$ is the slope of the quasi-Fermi level extracted from a linear regression in the inner $z\in[-10, 10]$ nm range of the 30 nm long channel.

The mobility and contact resistance obtained that way are plotted in Fig. \ref{figR10x10qfl}. They are in very good agreement with those extracted with the transmission line methodology of paragraph \ref{subsectionMethodology} in the strong inversion regime, where the quasi-Fermi level drops quasi-linearly under the gate. The contact resistance is, therefore, indeed dominated by the spacers.\cite{Rideau14} The mobility also compares well with that extracted from the $R(L)$ data on ``homogeneous'' nanowire channels without spacers and bulk source/drain contacts (the gate then runs across the whole simulation box -- see Ref. \onlinecite{Niquet14}). The differences at low $\langle n_{\rm 2d}\rangle_{\rm ch}$ are due to the source/channel and channel/drain junctions, whose finite depletion widths result in significant variations of the carrier density near both ends of the channel (see Fig. \ref{figDensity}). As a matter of fact, homogeneous nanowire channels without such junctions are better suited to the calculation of the long-channel mobility at low carrier density.\cite{Niquet14} The calculated mobilities are close to the experimental data of Ref. \onlinecite{Coquand13b} whatever the methodology.


The analysis of the quasi-Fermi level provides an alternative to the $R(L)$ methodology of paragraph \ref{subsectionMethodology} (faster but often less accurate at low densities). Although Fig. \ref{figQFL} is consistent with a drift-diffusion model using different mobilities in the channel, spacers and bulk source/drain, we point out that Eq. (\ref{eqDD}) is not able to capture the ballistic component of the contact resistance on physical grounds. $\bar{R}_{\rm 0}$ might indeed be lumped in the mobility under the spacers, but that mobility would hardly be transferable to other device designs.

In the next section, we discuss the different (ballistic and diffusive) contributions to the contact resistance.

\section{Analysis of the contact resistance and models}
\label{sectionBreakdown}

In this section, we break down the contact resistance into ballistic and diffusive contributions and discuss simple models for the NEGF data.

\subsection{Breakdown of the contact resistance into ballistic and diffusive contributions}

\begin{figure}
\includegraphics[width=0.98\columnwidth]{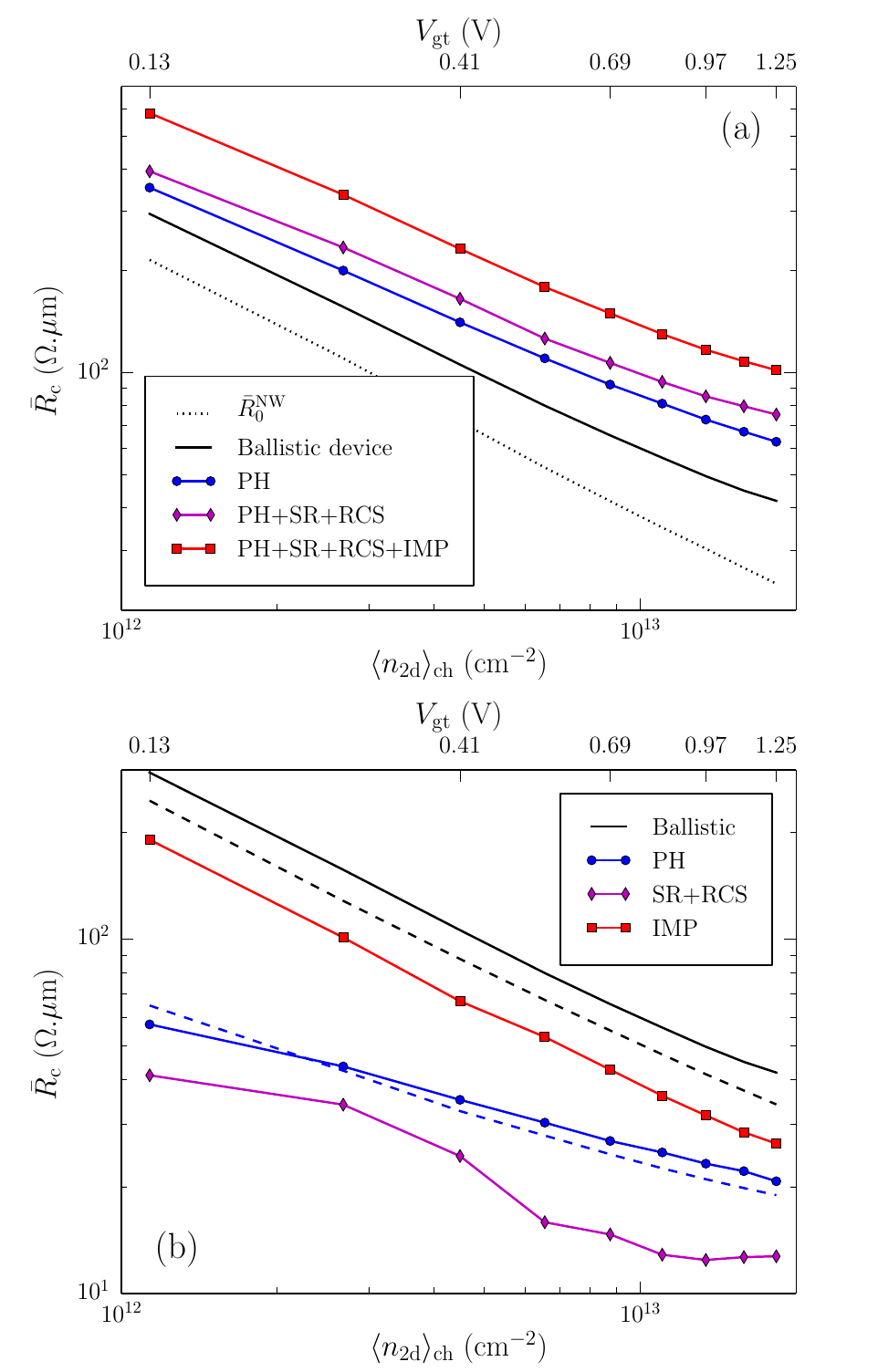}
\caption{(a) Contact resistance $\bar{R}_{\rm c}$ as a function of the carrier density in a $W=10\times H=10$ nm trigate device with the ``Reference $1.5\times 10^{20}$ cm$^{-3}$'' doping profile of Fig. \ref{figDoping}. Different scattering mechanisms have been included (see text). (b) The ballistic, phonon (PH), surface roughness (SR+RCS), and impurity (IMP) scattering contributions to the contact resistance, obtained as differences between the data of Fig. \ref{figContrib}a. The dashed lines are simple models for the ballistic resistance and phonons contribution (see text).}
\label{figContrib}
\end{figure}

In order to disentangle the main contributions to the contact resistance, we have computed the following series of $W=10\times H=10$ nm devices:
\begin{enumerate}
 \item The ``Ballistic'' device without scattering (neither phonons nor SR, RCS and impurities). In order to preserve the electrostatics, the point-like dopants in the source and drain are replaced by the background distribution of Fig. \ref{figDoping}, and the RCS charges by a uniform background at the SiO$_2$/HfO$_2$ interface. As expected, the resistance $\bar{R}_0$ of this device is independent on the device length in the $L\ge 30$ nm range. It is plotted in Fig. \ref{figR10x10}.
 \item Same as 1, with phonons enabled (PH).
 \item Same as 2, with surface roughness and RCS enabled (PH+SR+RCS).
 \item Same as 3, with Coulomb scattering enabled in the source and drain (PH+SR+RCS+IMP, i.e. full device with point-like dopants).
\end{enumerate}
The different contact resistances are plotted as a function of $\langle n_{\rm 2d}\rangle_{\rm ch}$ in Fig. \ref{figContrib}a, while the contributions from phonons (PH), surface roughness and remote Coulomb (SR+RCS), and impurity (IMP) scattering, defined as the differences between the contact resistance of two successive devices, are plotted in Fig. \ref{figContrib}b. As before, the data were averaged over different realizations of the disorders.

The ballistic resistance is the dominant contribution to $\bar{R}_c$, but represents only 40\% to 50\% of the total contact resistance. The main diffusive component is impurity scattering, followed by phonons, then SR+RCS scattering. As expected, the ratio impurity/phonon scattering decreases with increasing carrier density due to the screening of the ionized impurities by the electron gas under the spacers and gate. The SR+RCS contribution is actually dominated by RCS scattering near both ends of the channel at low densities, and by SR scattering under the spacers at large densities. It is, however, difficult to disentangle these two smaller contributions. Yet SR scattering clearly makes a minor contribution to $\bar{R}_c$, while it is a dominant mechanism in the channel. Actually, most of the contacts undergo volume accumulation rather than surface inversion, so that the carriers do not probe much the surface of the spacers. The distribution of carriers under the spacers will be discussed in more detail in the next paragraph.

\subsection{Models and discussion}

In order to get a deeper understanding of Fig. \ref{figContrib}, we have computed the ballistic resistance $\bar{R}_0^{\rm NW}(n_{\rm 2d})$ of a homogeneous (infinitely long) $10\times 10$ nm nanowire channel (dotted line in Fig. \ref{figContrib}a).\cite{NoteBall} $\bar{R}_0^{\rm NW}(n_{\rm 2d})$ is significantly lower than the ballistic resistance $\bar{R}_0$ of the whole device. The cross section and carrier density are, actually, very inhomogeneous in the source/drain and under the spacers. In the simplest ``top of the barrier'' model,\cite{Natori94,Rahman03} the ballistic resistance of such an inhomogeneous device shall be the ballistic resistance of the homogeneous conductor built from the cross section at the top of the 1D conduction band edge profile $E_c(z)$. At low bias, the top of the barrier is very close to the point $z=z_{\rm min}$ where the carrier density $n_{\rm 2d}(z)\equiv n_{\rm 2d}^{\rm min}$ reaches its minimum. We have, therefore, plotted $\bar{R}_0^{\rm NW}(n_{\rm 2d}^{\rm min})$ as a dashed black line on Fig. \ref{figContrib}b. This simple estimate matches the resistance of the ballistic device much better. The point $z=z_{\rm min}$ is under the gate in the subthreshold regime (Fig. \ref{figDensity}), but moves under the spacers in the ON regime. The ballistic resistance is then more dependent on the design of the spacers than on the design of the channel, and truly appears in this respect as a ``contact'' resistance. 

We have also attempted to recompute the phonon contribution from a simple local mobility model:
\begin{equation}
\bar{R}_{\rm PH}=\int_{\rm device}\frac{dz}{n_{\rm 2d}(z)\mu^{\rm PH}(z)e}-\frac{L}{\langle n_{\rm 2d}\rangle_{\rm ch}\mu^{\rm PH}_{\rm ch}e}\,,
\end{equation}
where $n_{\rm 2d}(z)$ is the carrier density, $\mu^{\rm PH}(z)$ the phonon-limited mobility at $z$, and $\mu^{\rm PH}_{\rm ch}$ the inversion layer mobility in the channel. The first term is resistance of the whole device, while the second term is the resistance of an ideal channel with homogeneous carrier density $\langle n_{\rm 2d}\rangle_{\rm ch}$. We can further approximate this expression as follows:
\begin{align}
\bar{R}_{\rm PH}&\sim\frac{1}{\mu^{\rm PH}_{\rm ch}e}\left[\int_{\rm device}\frac{dz}{n_{\rm 2d}(z)}-\frac{L}{\langle n_{\rm 2d}\rangle_{\rm ch}}\right] \nonumber \\
&\sim\frac{L_{\rm ref}}{\left\langle\Delta n_{\rm 2d}^{-1}\right\rangle^{-1}\mu^{\rm PH}_{\rm ch}e}\,,
\label{eqRmodel}
\end{align}
where
\begin{equation}
\left\langle\Delta n_{\rm 2d}^{-1}\right\rangle=\frac{1}{L_{\rm ref}}\left[\int_{\rm device}\frac{dz}{n_{\rm 2d}(z)}-\frac{L}{\langle n_{\rm 2d}\rangle_{\rm ch}}\right]\,,
\label{eqDn2d}
\end{equation}
and $L_{\rm ref}$ is an (arbitrary) reference length. $\langle\Delta n_{\rm 2d}^{-1}\rangle$ is the difference between the inverse carrier densities integrated in the whole device and in the ideal channel, normalized to $L_{\rm ref}$. We choose $L_{\rm ref}=2L_{\rm sp}=12$ nm, so that Eq. (\ref{eqRmodel}) appears as an effective spacer resistance with an effective spacer carrier density $\langle n_{\rm 2d}\rangle_{\rm sp}\equiv\left\langle\Delta n_{\rm 2d}^{-1}\right\rangle^{-1}$.  The above approximation on the local mobility is justified by the fact that the main contributions to $\langle\Delta n_{\rm 2d}^{-1}\rangle$ and $\bar{R}_{\rm PH}$ are collected under the spacers (where $1/n_{\rm 2d}(z)$ reaches its maximum in the strong inversion regime), and near the entrance/exit of the channel (where the differences between $n_{\rm 2d}(z)$ and $\langle n_{\rm 2d}\rangle_{\rm ch}$ are the largest). In these regions, $\mu^{\rm PH}(z)$ can be replaced with $\mu^{\rm PH}_{\rm ch}$, as $\mu^{\rm PH}$ shows a rather weak dependence on the carrier density. This model, plotted as a dashed blue line in Fig. \ref{figContrib}, reproduces the NEGF data very well. It also reproduces the main trends, but is not as accurate for the surface roughness and impurity contributions, whose mobilities show stronger dependences on the carrier density. The carriers indeed move from a ``bulk-like'' distribution in the source to ``surface-inversion-like'' distribution in the channel (Fig. \ref{figDensity}), so that neither the inversion layer nor the bulk mobilities hold under the spacers. In particular, the SR-limited inversion layer mobility $\mu^{\rm SR}_{\rm ch}$ overestimates the surface roughness contribution to $\bar{R}_c$. Indeed, $n_{\rm 2d}(z)$ can be pretty large near the source/spacer junction, but the carriers are distributed in the bulk of the nanowire, and are, therefore much less scattered by surface roughness than expected from $\mu^{\rm SR}_{\rm ch}$. 

\begin{figure}
\includegraphics[width=0.98\columnwidth]{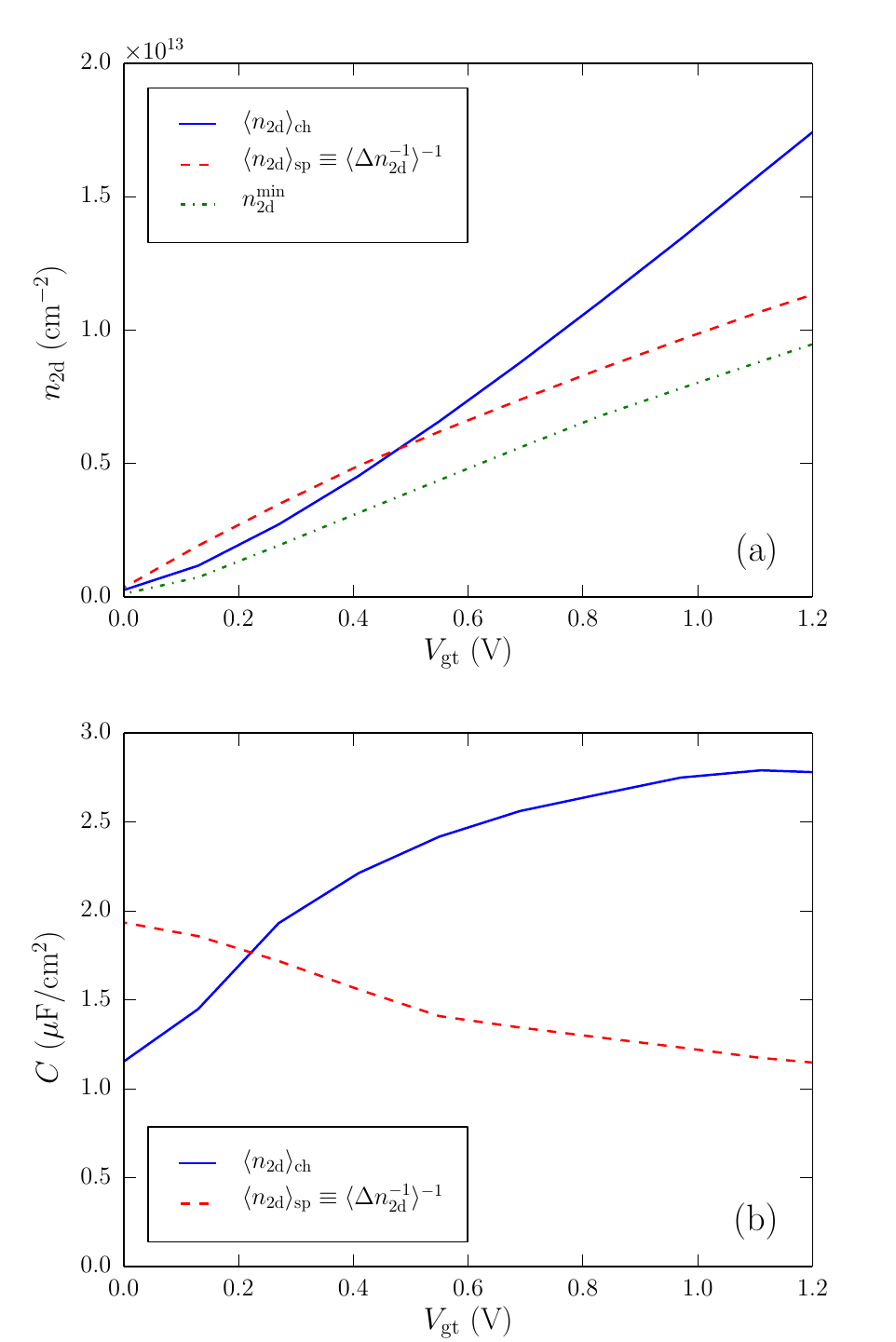}
\caption{(a) Average carrier density $\langle n_{\rm 2d}\rangle_{\rm ch}$ in the channel and $\langle n_{\rm 2d}\rangle_{\rm sp}$ under spacers, and minimal density $n_{\rm 2d}^{\rm min}$ in the device as a function of the gate overdrive $V_{\rm gt}$, in a $W=10\times H=10$ nm trigate device with the ``Reference $1.5\times 10^{20}$ cm$^{-3}$'' doping profile of Fig. \ref{figDoping}. The point where the density reaches its minimum $n_{\rm 2d}^{\rm min}$ is under the spacers when $V_{\rm gt}>0.4$ V. (b) Gate-channel and gate spacers capacitances. The data were averaged over five different devices.}
\label{figCV}
\end{figure}

As discussed above, the conductivity of the contacts (as well as the ballistic conductance in the strong inversion regime) are essentially controlled by the density of carriers under the spacers. The latter can be adequately characterized by $\langle n_{\rm 2d}\rangle_{\rm sp}\equiv\left\langle\Delta n_{\rm 2d}^{-1}\right\rangle^{-1}$. By definition [Eq. (\ref{eqDn2d})], the spacers are depleted (as regards transport) with respect to the channel when $\langle n_{\rm 2d}\rangle_{\rm sp}\lesssim\langle n_{\rm 2d}\rangle_{\rm ch}$. $\langle n_{\rm 2d}\rangle_{\rm ch}$ and $\langle n_{\rm 2d}\rangle_{\rm sp}$ exhibit different dependences on the gate voltage (Figs. \ref{figDensity} and \ref{figCV}). \corr{The channel is indeed well controlled by the gate, the gate to channel capacitance $C_{\rm ch}=d\langle n_{\rm 2d}\rangle_{\rm ch}/dV_{\rm gt}$ reaching a large fraction of the oxide capacitance $C_{\rm ox}$. On the contrary, the spacers are doped but are poorly controlled by the gate (because they do not overlap). The gate to spacers capacitance $C_{\rm sp}=d\langle n_{\rm 2d}\rangle_{\rm sp}/dV_{\rm gt}$ is actually much lower than $C_{\rm ch}$ (and weakly dependent on $V_{\rm gt}$). The spacers do not, therefore, accumulate much excess charge in the ON regime. In the present device, $\langle n_{\rm 2d}\rangle_{\rm ch}$ and $\langle n_{\rm 2d}\rangle_{\rm sp}$ cross at gate overdrive $V_{\rm gt}\simeq 0.5$ V. Above that gate voltage, the carrier density shows a significant dip under the spacers (with density $n_{\rm 2d}^{\rm min}$) that makes a significant contribution to the ballistic and diffusive resistances.}


To assess charge control in different devices, we can further approximate $\langle n_{\rm 2d}\rangle_{\rm ch}$ and $\langle n_{\rm 2d}\rangle_{\rm sp}$ by:
\begin{align}
\langle n_{\rm 2d}\rangle_{\rm ch}&\simeq C_{\rm ox}\left(V_{\rm g}-\hat{V}_{\rm t}\right) \nonumber \\
\langle n_{\rm 2d}\rangle_{\rm sp}&\simeq C_{\rm sp}\left(V_{\rm g}-\hat{V}_{\rm t}^\prime\right)\,,
\label{eqnchnsp}
\end{align}
where $C_{\rm ox}\approx 2.7$ $\mu$F/cm$^2$ for all devices considered in this work, and $\hat{V}_{\rm t}$ and $\hat{V}_{\rm t}^\prime$ are threshold voltages for the channel and spacer, respectively.\cite{NoteVt} There is, therefore, a linear relation between $\langle n_{\rm 2d}\rangle_{\rm sp}$ and $\langle n_{\rm 2d}\rangle_{\rm ch}$:
\begin{equation}
\langle n_{\rm 2d}\rangle_{\rm sp}-n_0=\alpha\left(\langle n_{\rm 2d}\rangle_{\rm ch}-n_0\right)\,,
\label{eqnsp}
\end{equation}
where $\alpha=C_{\rm sp}/C_{\rm ox}<1$ and $n_0=C_{\rm sp}C_{\rm ox}(\hat{V}_{\rm t}-\hat{V}_{\rm t}^\prime)/(C_{\rm ox}-C_{\rm sp})$ is the crossover density. $\alpha$ and $n_0$ are characteristic of a design, but do not depend on the channel length (at least in the absence of short-channel effects). As an example, $\alpha=0.47$ and $n_0=6.1\times 10^{12}$ cm$^{-2}$ in the $10\times 10$ nm device of Fig. \ref{figCV}. Data for other devices will be given in the next paragraphs. In the strong inversion regime, $\bar{R}_{\rm c}$ is approximately proportional to $\langle n_{\rm 2d}\rangle_{\rm sp}^{-1}$. It can therefore be written $\bar{R}_{\rm c}\sim 2L_{\rm sp}/(\langle n_{\rm 2d}\rangle_{\rm sp}\mu_{\rm sp}e)$, where $\mu_{\rm sp}$ is an ``effective'' spacer mobility (typically in the $55-65$ cm$^2$/V/s range). We would like to stress, though, that $\mu_{\rm sp}$ embeds the ballistic resistance and shall not, therefore, be interpreted as a diffusive mobility transferable to other spacer lengths.


To conclude, we would like to remind that the present NEGF calculations miss the contribution from the metal/semiconductor contact resistance. The latter is expected to be weakly dependent on $V_{\rm gt}$, and shall therefore appear as a rigid shift $\bar{R}^\prime_{\rm c}=R^\prime_{\rm c}W_{\rm eff}$, where $R^\prime_{\rm c}$ is independent of $V_{\rm gt}$ and of $W_{\rm eff}$ (as long as the metal/semiconductor contact remains the same whatever the design of the channel).

\section{Importance of the design of the near spacer region}
\label{sectionDesign}

As discussed above, the design of the spacers can have a great impact on the contact resistance. We investigate in this section the effect of the cross section of the nanowire (for a given doping profile), then the effect of the doping profile (at constant cross section), and finally the effect of charges trapped in the spacer material.

\subsection{Influence of the cross section}
\label{subsectionSection}

\begin{figure*}
\includegraphics[width=0.96\textwidth]{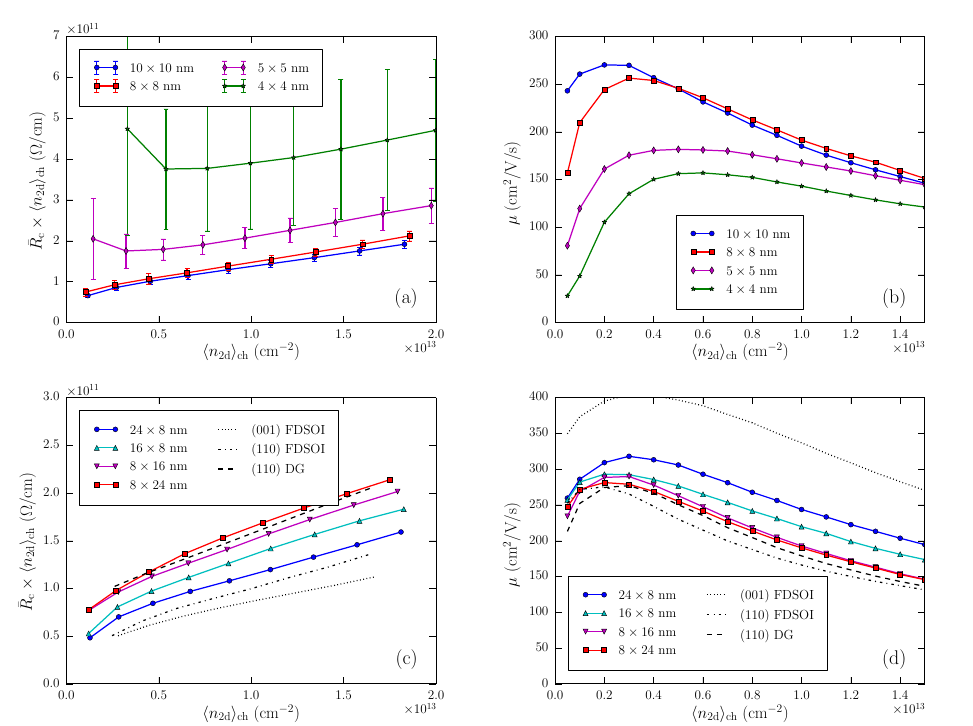}
\caption{Contact resistance $\bar{R}_{\rm c}\times\langle n_{\rm 2d}\rangle_{\rm ch}$ (a, c) and channel mobility $\mu$ (b, d) as a function of the carrier density $\langle n_{\rm 2d}\rangle_{\rm ch}$ in the channel, for devices with different cross sections $W\times H$. The doping profile is ``Reference $1.5\times 10^{20}$ cm$^{-3}$'' from Fig. \ref{figDoping}. (a, b) are ``square'' trigate devices ($W=H$). The standard deviation on the contact resistance, representative of local variability, is represented by the error bars ($\pm 1\sigma$). (c, d) are``rectangular'' trigate and FinFET devices with either $W=8$ nm or $H=8$ nm. The mobility and contact resistances of planar $(001)/(110)$ FDSOI devices and symmetric $(110)$ double gate devices (DG) are also reported for comparison.\cite{noteplanar}}
\label{figRcsection}
\end{figure*}

The contact resistance and channel mobility are plotted as a function of the carrier density $\langle n_{\rm 2d}\rangle_{\rm ch}$ for trigate and FinFET devices with different cross sections $W\times H$ in Fig. \ref{figRcsection}. They have been extracted from linear regressions on the $R(L)$ data on full devices and on homogeneous channels,\cite{Niquet14} respectively. In order to improve the readability of the figure, we have actually plotted $\bar{R}_{\rm c}\times\langle n_{\rm 2d}\rangle_{\rm ch}$ as a function of $\langle n_{\rm 2d}\rangle_{\rm ch}$ in Fig. \ref{figRcsection}a,c. Fig. \ref{figRcsection}a,b display the results for ``square'' trigate devices with side $W=H$ ranging from 4 to 10 nm, and Fig. \ref{figRcsection}c,d the results for horizontal trigate and vertical FinFET devices with either $W=8$ nm or $H=8$ nm. The doping profile is ``Reference $1.5\times 10^{20}$ cm$^{-3}$'' from Fig. \ref{figDoping}. The standard deviation on $\bar{R}_{\rm c}$ ($\pm 1\sigma$) computed on a set of eight different devices is also reported for square trigate devices as a measure of local variability.

The carrier mobility in the channel of square trigates decreases when the cross section is reduced due to the enhancement of scattering by structural confinement. The picture is, however, very dependent on the carrier density. At high density, inversion mostly takes place at the surface of the nanowire (except in the $4\times 4$ nm device). Hence the current flows on the top and two side facets, and transport along each facet is mainly limited by SR scattering on that facet. The mobility is, therefore, little dependent on the cross section. The carriers, however, flow closer to the axis of the nanowire when decreasing $\langle n_{\rm 2d}\rangle_{\rm ch}$ and $W$, and get eventually scattered by the other facets of the nanowire. This explains the strong dependence of the mobility on the cross section at intermediate carrier densities. At small carrier densities, the transport becomes limited by Remote Coulomb scattering.

In square trigate devices, the normalized contact resistance $\bar{R}_{\rm c}$ also tends to increase with decreasing cross section, especially below $W=H=8$ nm. In the strong inversion regime, $\bar{R}_{\rm c}$ is about three times larger in a $4\times 4$ nm than in a $10\times 10$ nm device, hence the {\it denormalized} contact resistance $\bar{R}_{\rm c}/W_{\rm eff}$ is almost 7 times larger. The same analysis as in section \ref{sectionBreakdown} shows that the ballistic and diffusive components of $\bar{R}_{\rm c}$ are both enhanced by confinement in the $4\times 4$ nm device. The main contributors at high inversion are now impurity scattering, then surface roughness scattering, the ballistic resistance, and finally electron-phonon scattering. The increase of diffusive components partly results, as for the channel mobility, from the enhancement of scattering by confinement. Yet the degradation of $\bar{R}_{\rm c}$ is mostly due to the decrease of the carrier density under the spacers. The crossover density indeed decreases when reducing cross section (down to $n_0\simeq 5.4\times 10^{11}$ cm$^{-2}$ in the $4\times 4$ nm trigate), so that the spacers become a strong bottleneck already at weak inversion densities. This results from the decrease of the number of dopants under the spacers and from the increase of the surface to volume ratio. Indeed, on the one hand, the linear density of carriers and conductivity under the spacers are expected to be proportional (in the simplest approximation) to the number of dopants per unit length, $n_d\propto WH$. On the other hand, the spacers feed a channel with total width $W_{\rm eff}=W+2H$. Therefore, the normalized contact resistance of a square trigate is expected to behave as ${\bar R}_{\rm c}\propto (W+2H)/(WH)\propto 1/W$. As a consequence, the contact resistance can be as large as 75\% of the total device resistance in a 30 nm long, $4\times 4$ nm trigate at gate overdrive $V_{\rm gt}\simeq 0.7$ V (see Table \ref{tableVariability}).

The device to device variability also increases a lot when reducing cross section. As the average number of dopants under each spacer decreases from $\simeq 13$ in the $10\times 10$ nm device to only $\simeq 2$ in the $4\times 4$ nm device, the fluctuations of the number and position of these dopants have increasing impact on the electrostatics and transport properties of the transistors.\cite{Seoane09,Martinez12,Markov12}

\begin{table}
\begin{tabular}{l|rr|rr|r}
\toprule
$W\times H$ & $\langle N_{\rm imp}\rangle$ & $\displaystyle\frac{\sigma_{\rm imp}}{\langle N_{\rm imp}\rangle}$ & $n_0$ (cm$^{-2}$) & $\alpha$ & $\displaystyle\frac{R_{\rm c}}{R(30\text{ nm})}$ \\
\hline
$4\times 4$ nm   & 1.9 & 0.74 & $5.4\times 10^{11}$ & 0.36 & 0.75 \\
$5\times 5$ nm   & 3.1 & 0.58 & $1.8\times 10^{12}$ & 0.39 & 0.67 \\
$8\times 8$ nm   & 8.1 & 0.36 & $4.9\times 10^{12}$ & 0.43 & 0.61 \\
$10\times 10$ nm & 12.7 & 0.28 & $6.1\times 10^{12}$ & 0.47 & 0.57 \\
\hline
$8\times 16$ nm  & 16.3 & 0.24 & $5.0\times 10^{12}$ & 0.46 & 0.59 \\ 
$8\times 24$ nm  & 24.6 & 0.19 & $4.7\times 10^{12}$ & 0.48 & 0.62 \\ 
$16\times 8$ nm  & 16.3 & 0.24 & $7.2\times 10^{12}$ & 0.46 & 0.60 \\ 
$24\times 8$ nm  & 24.6 & 0.19 & $9.8\times 10^{12}$ & 0.49 & 0.58 \\ 
\botrule
\end{tabular}
\caption{Average number of impurities $\langle N_{\rm imp}\rangle$ under each spacer and normalized standard deviation $\sigma_{\rm imp}/\langle N_{\rm imp}\rangle$ in trigate devices with the ``Reference $1.5\times 10^{20}$ cm$^{-3}$'' doping profile of Fig. \ref{figDoping}. Crossover density $\langle n_0\rangle$, ratio $\alpha=C_{\rm sp}/C_{\rm ox}$, and ratio between the average contact resistance $R_{\rm c}$ and the total resistance $R(L=30\text{ nm})$ of 30 nm long devices, at gate overdrive $V_{\rm gt}=0.7$ V.}
\label{tableVariability}
\end{table}


\begin{figure}
\includegraphics[width=0.98\columnwidth]{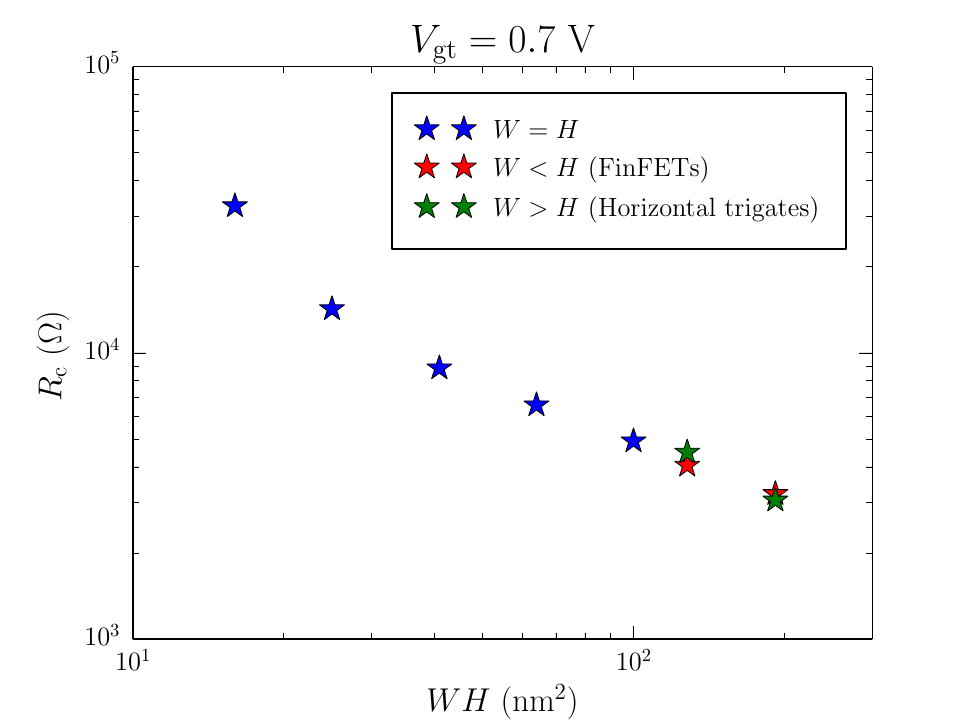}
\caption{Contact resistance $R_{\rm c}$ (per wire) as a function of the cross-sectional area $S=WH$ in different kinds of devices (square and horizontal trigates, vertical FinFETs) at gate overdrive $V_{\rm gt}=0.7$ V.}
\label{FigRcS}
\end{figure}

The contact resistance and mobility in rectangular trigate and FinFET devices (Figs. \ref{figRcsection}c, d) also show sizable dependence on the cross section. Actually, vertical FinFET devices with $H\gg W$ are expected to reach the same contact resistance and mobility $\mu_{110}$ as planar $(110)$, symmetric double-gate (DG) devices,\cite{noteplanar} while horizontal trigate devices with $W\gg H$ are expected to reach the same contact resistance and mobility $\mu_{001}$ as planar $(001)$ FDSOI devices.\cite{noteplanar} $\mu_{110}$ is smaller than $\mu_{001}$ due to the stronger surface roughness and heavier transport mass along $(110)$ facets [the ground state valley is the $Z$ valley with transport mass $m^*=0.19\,m_0$ in $(001)$ films and the heavier $X,Y$ valleys with transport mass $m^*\approx 0.32\,m_0$ in $(110)$ films]. The mobility is therefore larger in trigate devices with $W>H$ than in FinFET devices with $H>W$. As an example, the $W=24\times H=8$ nm trigate shows 30\% larger mobility than the $W=8\times H=24$ nm FinFET at high inversion density ($\langle n_{\rm 2d}\rangle_{\rm ch}=10^{13}$ cm$^{-2}$). It also shows 30\% lower contact resistance, but this performance can hardly be related with the better channel mobilities. As a matter of fact, the contact resistance of planar $(001)$ and $(110)$ FDSOI devices are quite similar while their mobilities are very different (primarily because the spacers do not undergo as strong surface inversion as the channel). Likewise, the denormalized contact resistance $R_{\rm c}=\bar{R}_{\rm c}/W_{\rm eff}$ of the $W=8{\ \rm nm}\times H=a$ FinFET and of the $W=a\times H=8$ nm trigate are very close. As pointed out for square trigates, and confirmed by Fig. \ref{FigRcS}, $R_{\rm c}$ is almost inversely proportional to the cross sectional area $S=WH$ of the spacers in this range of dimensions (except for the smallest devices). However, the spacers of the vertical FinFET feed a much larger channel ($W_{\rm eff}=2a+8$ nm) than those of the horizontal trigate ($W_{\rm eff}=a+16$ nm). This explains why the normalized contact resistance $\bar{R}_{\rm c}$ of the vertical FinFETs shows a weak dependence on $H$, and remains very close to the reference $(110)$ double gate device, while the normalized contact resistance of horizontal trigate devices rapidly improves with increasing $W$. In general, devices that maximize effective channel width for a given cross section will show larger normalized contact resistances.


\subsection{Influence of the doping profile}
\label{sectionDoping}

\begin{table*}
\begin{tabular}{l|rrrr|r|rr}
\toprule
Name & $N_d^0$ (cm$^{-3}$) & $\lambda_1$ (nm) & $z_2-z_c$ (nm) & $\lambda_2$ (nm) & $\langle N_{\rm imp}\rangle$ & $n_0$ (cm$^{-2}$) & $\alpha$ \\
\hline
Reference $1.5\times 10^{20}$ cm$^{-3}$ & $1.5\times 10^{20}$ & $3.0$ & $0.0$ & $+\infty$ & 12.7 & $6.1\times 10^{12}$ & 0.47 \\
Reference   $3\times 10^{20}$ cm$^{-3}$ &   $3\times 10^{20}$ & $3.0$ & $0.0$ & $+\infty$ & 25.4 & $1.3\times 10^{13}$ & 0.61 \\
Strong underlap & $3\times 10^{20}$ & $4.0$ & $-2.0$ & $1.5$ & 28.4 & $1.4\times 10^{13}$ & 0.67 \\
Small underlap  & $3\times 10^{20}$ & $4.0$ &  $0.0$ & $1.5$ & 32.4 & $2.1\times 10^{13}$ & 0.64 \\
Small overlap   & $3\times 10^{20}$ & $4.0$ &  $4.0$ & $1.5$ & 34.2 & $2.3\times 10^{13}$ & 0.49 \\
\botrule
\end{tabular}
\caption{Parameters for the different doping profiles of Fig. \ref{figDoping} [See Eq. (\ref{eqNd})]. The average number of impurities $\langle N_{\rm imp}\rangle$ under the spacers of a $W=10\times H=10$ nm trigate device, the crossover density $\langle n_0\rangle$, and the ratio $\alpha=C_{\rm sp}/C_{\rm ox}$ are also given.}
\label{tableDoping}
\end{table*}

In this paragraph, we discuss the impact of the doping profile on the contact resistance in a $W=10\times H=10$ nm trigate device.

We have considered the different doping profiles plotted in Fig. \ref{figDoping}. On the source side, the ionized impurity concentration $N_d(z)$ reads:
\begin{equation}
N_d(z)=N_d^0\times\frac{1}{1+10^{(z-z_1)/\lambda_1}}\times\frac{1}{1+10^{(z-z_2)/\lambda_2}}\,,
\label{eqNd}
\end{equation}
where $N_d^0$ is the dopant concentration deep in the source ($N_d^0=1.5\times 10^{20}$ cm$^{-3}$ or $N_d^0=3\times 10^{20}$ cm$^{-3}$). \corr{$N_d(z)$ is approximately equal to $N_d^0$ when $z<z_1$, then decreases by almost one order of magnitude every $\lambda_1$ when $z_1<z<z_2$, and one order of magnitude every $(\lambda_1^{-1}+\lambda_2^{-1})^{-1}$ when $z>z_2$.} In the following, $z_1=z_c-6$ nm is the entrance of the source spacer ($z_c$ being the entrance of the channel), while $z_2$ can be varied around $z_c$ (see Table \ref{tableDoping}). The ``Reference $1.5\times 10^{20}$ cm$^{-3}$'' and ``Reference $3\times 10^{20}$ cm$^{-3}$'' profiles feature one single decay length $\lambda_1=3$ nm/decade ($\lambda_2\to\infty$), while the other profiles feature two distinct decay lengths. The ``small underlap'' profile is an optimal distribution resulting from a process simulation. The doping profiles are symmetric under the drain.

\begin{figure}
\includegraphics[width=0.98\columnwidth]{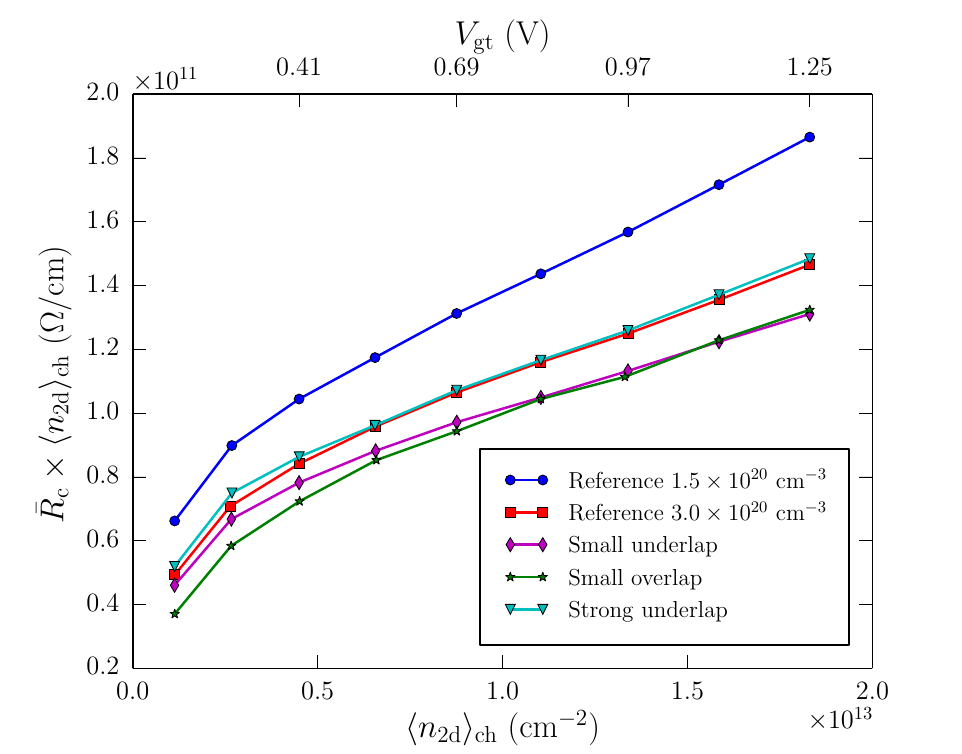}
\caption{Contact resistance $\bar{R}_{\rm c}\times\langle n_{\rm 2d}\rangle_{\rm ch}$ in a $W=10\times H=10$ nm trigate device as a function of the gate overdrive (upper axis)/carrier density in the channel (lower axis), for the different doping profiles of Fig. \ref{figDoping}.}
\label{figRcdoping}
\end{figure}

The contact resistances $\bar{R}_{\rm c}$ computed in the $W=10\times H=10$ nm trigate device are plotted as a function of gate overdrive/carrier density in Fig. \ref{figRcdoping}. The devices with $N_d^0=3\times 10^{20}$ cm$^{-3}$ all show lower resistances than the devices with $N_d^0=1.5\times 10^{20}$ cm$^{-3}$. Indeed, larger $N_d^0$'s tend to increase the carrier density in the spacers [larger $n_0$ in Eq. (\ref{eqnsp})], hence to decrease the contact resistance. As a matter of fact, $n_0$ is greater than $10^{13}$ cm$^{-2}$ in all $N_d^0=3\times 10^{20}$ cm$^{-3}$ devices. 

In general, slower decay under the spacer also decreases the contact resistance, as shown by the comparison between the devices with $N_d^0=3\times 10^{20}$ cm$^{-3}$. The effect is not spectacular though, as in a $10\times 10$ nm channel the tail of impurities near the gate only contains on average one donor per nanometer at $N_d(z)=10^{19}$ cm$^{-3}$ -- the number of dopants at the entrance/exit of the channel remains therefore very low. 

The trends evidenced in Fig. \ref{figRcdoping} have been confirmed on a $W=8\times H=16$ nm FinFET device. We would like to point out that the above doping profiles have little impact on the carrier mobilities in the channel, and on the threshold voltage $V_{\rm t}$ at small $V_{\rm ds}$. Yet larger $N_d^0$ and slower decay under the spacers can increase drain-induced barrier lowering at large $V_{\rm ds}$ ; the problem of the contact resistances at high fields is, however, beyond the scope of the present work and will be discussed in an other paper. 

\subsection{Influence of charges trapped at the Si/Si$_3$N$_4$ interface}
\label{subsectionEmbedding}

Charges trapped in the channel at the Si/SiO$_2$ or SiO$_2$/HfO$_2$ interface are known to be strong scatterers, especially at low carrier densities. In this paragraph, we investigate the effects of charges trapped at the interface between the wire and the spacer material.

\begin{figure}
\includegraphics[width=0.98\columnwidth]{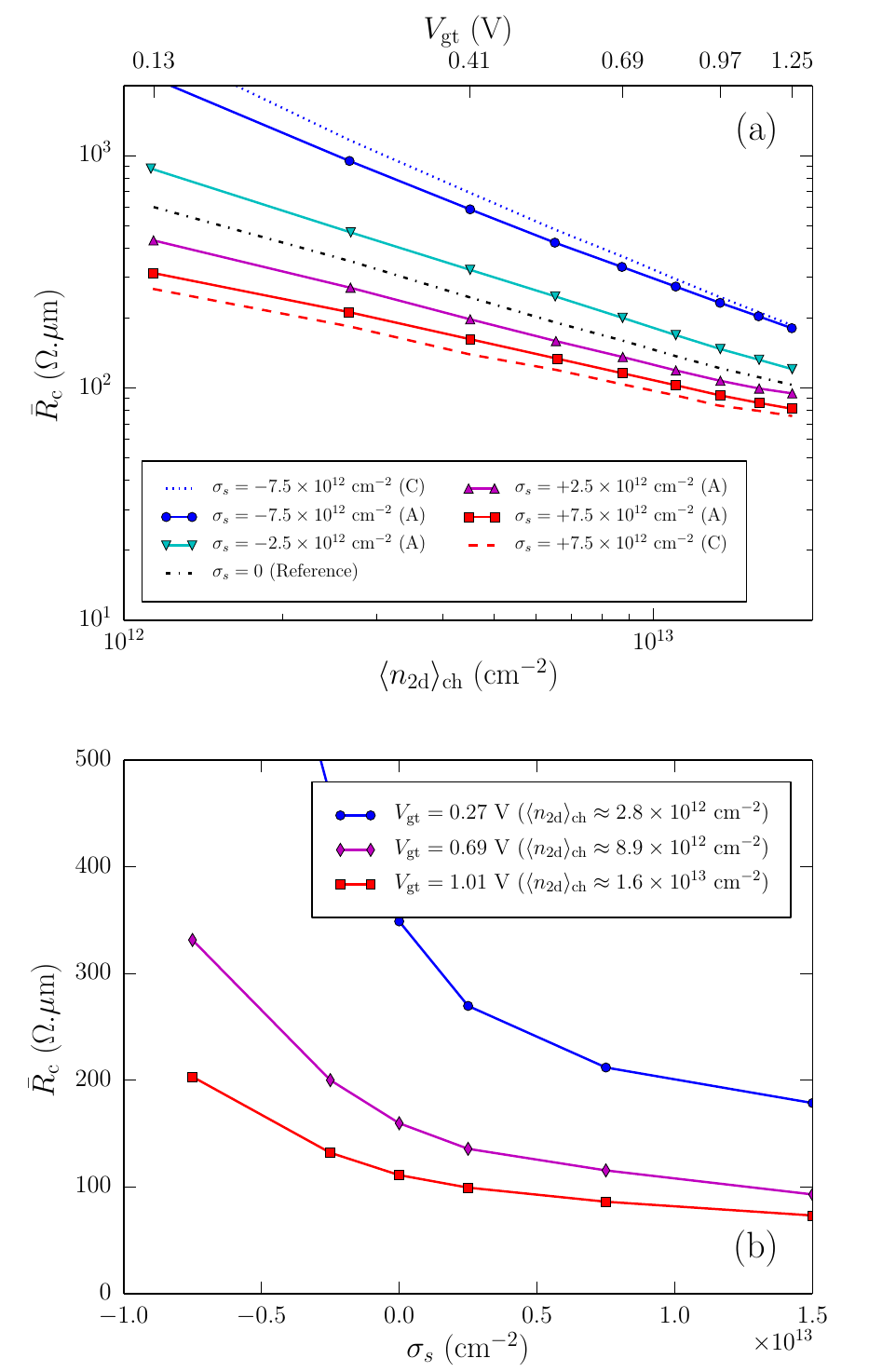}
\caption{(a) Contact resistance $\bar{R}_{\rm c}$ as a function of the gate overdrive (upper axis)/carrier density in the channel (lower axis), for different densities of trapped charges $\sigma_s$ at the Si/Si$_3$N$_4$ interface. The device is a $W=10\times H=10$ nm trigate with the ``Reference $1.5\times 10^{20}$ cm$^{-3}$'' doping profile of Fig. \ref{figDoping}. ``(A)'' means atomistic (point-like) charge distributions, and ``(C)'' means continuous background charge distributions at the Si/Si$_3$N$_4$ interface. (b) Contact resistance $\bar{R}_{\rm c}$ as a function of $\sigma_s$, for different gate overdrives/carrier density in the channel.}
\label{figRccharges}
\end{figure}

For that purpose, we have computed the contact resistances $\bar{R}_{\rm c}$ of a $W=10\times H=10$ nm trigate device, with different densities $\sigma_s$ of positive or negative charges randomly distributed at the Si/Si$_3$N$_4$ interface. $\bar{R}_{\rm c}$  is plotted in Fig. \ref{figRccharges}, as a function of gate overdrive for a given $\sigma_s$, and as a function of $\sigma_s$ for different gate overdrives. The threshold voltage $V_{\rm t}$ and the carrier mobility in the channel are little dependent on the density of traps.

The effects of positive and negative trap charges are opposite. Positive trap charges slightly decrease the contact resistance, whereas negative trap charges strongly increase it. The picture is symmetric for a $p$-type transistor: positive (resp. negative) trap charges increase (resp. decrease) the contact resistance. Also, the effect of the trap charges is almost the same whether they are distributed randomly at the Si/Si$_3$N$_4$ interface [curves labeled ``(A)'' on Fig. \ref{figRccharges}], or replaced by a continuous interface charge distribution [curves labeled ``(C)'' on Fig. \ref{figRccharges}]. Therefore, the variations of $\bar{R}_{\rm c}$ are essentially driven by electrostatics, and not by scattering (at variance with RCS for example). 

\begin{figure}
\includegraphics[width=0.98\columnwidth]{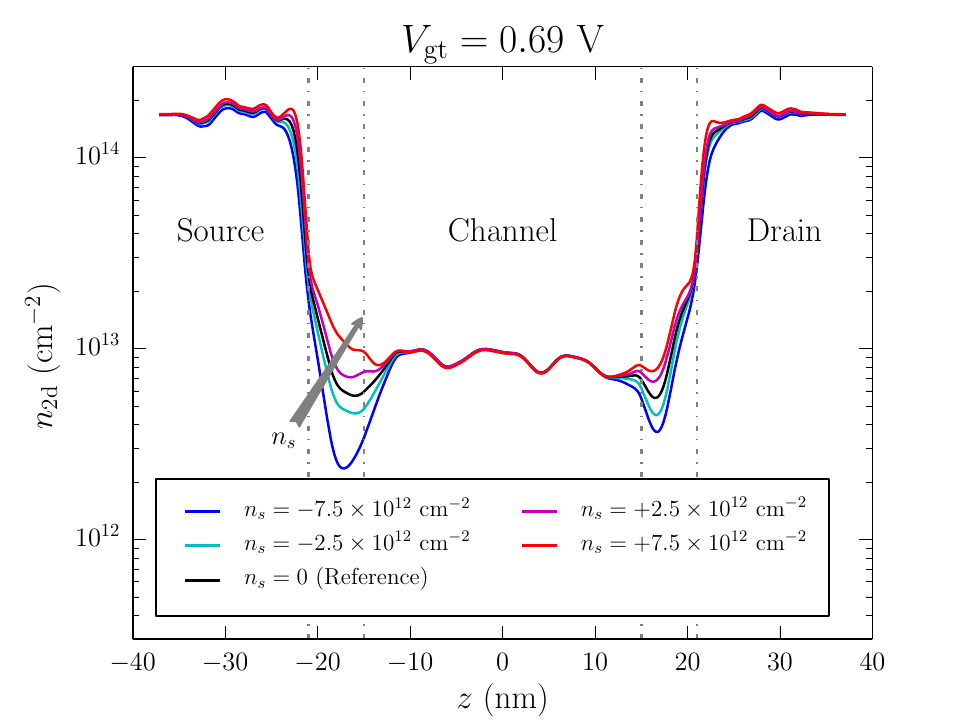}
\caption{Carrier density $n_{\rm 2d}$ in a 30 nm long, $W=10\times H=10$ nm trigate device with the ``Reference $1.5\times 10^{20}$ cm$^{-3}$'' doping profile of Fig. \ref{figDoping}, for different densities of trapped charges $\sigma_s$ at the Si/Si$_3$N$_4$ interface (gate overdrive $V_{\rm gt}=0.69$ V).}
\label{figDenscharges}
\end{figure}

As a matter of fact, interface traps behave as surface dopants, and can therefore enhance accumulation under the spacers ($n_0$ increases when $\sigma_s>0$), or deplete the wire ($n_0$ decreases when $\sigma_s<0$, see Fig. \ref{figDenscharges} and Table \ref{tableNs}). This either increases ($\sigma_s>0$) or decreases ($\sigma_s<0$) the conductivity under the spacers. The effect of these traps shall, therefore, be captured by any method that reproduces the correct electrostatics, including drift-diffusion models. 

\begin{table}
\begin{tabular}{r|r|rr}
\toprule
$\sigma_s$ (cm$^{-2}$) & $\langle Q_{\rm sp}\rangle$ ($e$) & $n_0$ (cm$^{-2}$) & $\alpha$ \\
\hline
$-7.5\times 10^{12}$ & $-0.3$ & $8.0\times 10^{11}$ & 0.38 \\
$-2.5\times 10^{12}$ &  $8.3$ & $3.5\times 10^{12}$ & 0.45 \\
     (Reference) $0$ & $12.7$ & $6.1\times 10^{12}$ & 0.47 \\
$ 2.5\times 10^{12}$ & $17.1$ & $7.9\times 10^{12}$ & 0.47 \\
$ 7.5\times 10^{12}$ & $25.7$ & $1.3\times 10^{13}$ & 0.48 \\
\botrule
\end{tabular}
\caption{Average dopant plus trapped charges $\langle Q_{\rm sp}\rangle$ under the spacers of a $W=10\times H=10$ nm trigate device with the ``Reference $1.5\times 10^{20}$ cm$^{-3}$'' doping profile of Fig. \ref{figDoping}, as a function of the density of trap charges $\sigma_s$ at the Si/Si$_3$N$_4$ interface. The crossover density $\langle n_0\rangle$ and the ratio $\alpha=C_{\rm sp}/C_{\rm ox}$ are also given.}
\label{tableNs}
\end{table}

The effects of surface traps increase with increasing surface to volume ratio under the spacers, i.e. with decreasing channel cross section, and with decreasing doping. As an example, $\bar{R}_{\rm c}(\sigma_s=-2.5\times 10^{12}\text{ cm}^{-2})-\bar{R}_{\rm c}(\sigma_s=0)$ is around twice larger in a $5\times 5$ nm than in a $10\times 10$ nm trigate at high inversion densities. Interestingly, the contact resistance remains almost the same if a 0.8 nm thick layer of SiO$_2$ is inserted between silicon and Si$_3$N$_4$, and the traps are moved at the SiO$_2$/Si$_3$N$_4$ interface. This results, again, from the electrostatic nature of the mechanism at play here, and can be contrasted with the behavior of RCS in the channel, whose scattering strength decreases exponentially with the thickness of the interfacial layer of SiO$_2$.\cite{Casse06} The effect of charges trapped in the bulk of the spacers is, as expected, weaker than the effect of charges trapped at the interface with silicon. Namely, $\bar{R}_{\rm c}(\sigma_s)\simeq\bar{R}_{\rm c}(\rho_s t_s)$ in the $10\times 10$ nm trigate of Fig. \ref{figRccharges}, where $\rho_s$ is the density of traps in the bulk of Si$_3$N$_4$, and $t_s=2.5$ nm is an effective thickness relating the effects of bulk charges to those of surface charges.
 
As a consequence of these results, a change of spacer material might improve the contact resistance of one kind of carrier (e.g., electrons) but degrade the contact resistance of the other one (e.g., holes) if the spacer and/or its interface with silicon gets charged differently.

\section{Comparison with experimental data}
\label{sectionExperimental}

We conclude this work by a comparison with experimental data on trigate devices with cross sections $W=14\times H=12$ nm and $W=40\times H=12$ nm fabricated at CEA/LETI.\cite{Coquand13} The gate stacks are made of 0.9 nm of SiO$_2$ and 2 nm of HfO$_2$ (effective oxide thickness ${\rm EOT}=1.3$ nm). The spacers are 9 nm long. The target doping in the source and drain is $N_d^0\simeq 3\times 10^{20}$ cm$^{-3}$; the doping profiles are modeled with Eq. (\ref{eqNd}) using $z_1-z_c=-9$ nm, $\lambda_1=6$ nm, and $z_2=z_c$ nm, $\lambda_2=1.5$ nm in the source. These values are characteristic of slightly underlapped devices.

\begin{figure}
\includegraphics[width=0.98\columnwidth]{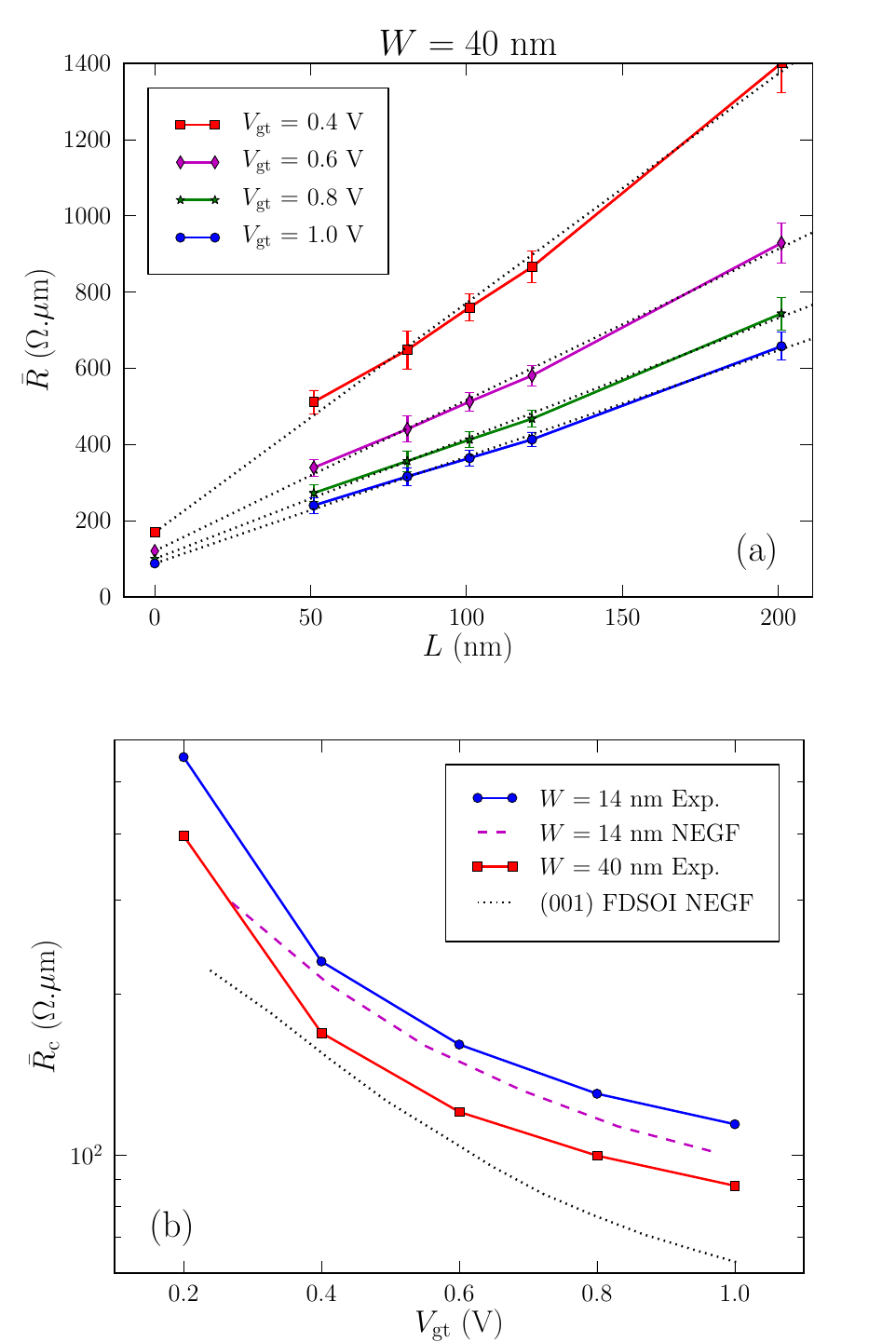}
\caption{(a) Total resistance measured as a function of gate length in $W=40\times H=12$ nm trigate devices, for different gate overdrives. (b) Measured contact resistances in $W=40\times H=12$ nm and $W=14\times H=12$ nm trigate devices, as a function of gate overdrive. They are compared with simulations for a $W=14\times H=12$ nm trigate and for a planar, 12 nm thick $(001)$ FDSOI device.}
\label{FigRLExp}
\end{figure}

The experimental contact resistances and mobilities are extracted with the same transmission line method as in the simulations in the $L=50-200$ nm range. The $R(L)$ data are averaged over 14 devices. The total resistance of the $W=40$ nm device is plotted as a function of gate length in Fig. \ref{FigRLExp}a (with $\pm 1\sigma$ error bars). \corr{It shows the expected linear behavior, although there is a small quadratic correction that might be due to the fact that long wires tend to thin in the middle.} The mobility in the $W=14$ nm device, measured as the slope of the $R(L)$ data, is close to the mobility in $10\times 10$ nm trigates (Fig. \ref{figR10x10qfl}), and is in excellent agreement with NEGF calculations. The contact resistances obtained from the linear regressions down to $L=0$ are plotted as a function of $V_{\rm gt}$ in Fig. \ref{FigRLExp}b, for both the $W=40$ nm and the $W=14$ nm devices. They are compared with NEGF simulations for the $W=14$ nm device and for a planar, 12 nm thick $(001)$ FDSOI device. 

\corr{The experimental data show the same $\propto V_{\rm gt}^{-\beta}$ dependence as the simulations. The experimental contact resistances for the $W=14$ nm device are, nonetheless, $10-20$ $\Omega$.$\mu$m larger than the simulations in the $V_{\rm gt}\ge 0.4$ V range. We remind, though, that the simulations miss the metal/semiconductor contact resistance, which shall actually be around 20 $\Omega.\mu$m.\cite{noteContact,Ohuchi12} The experimental data for the $W=40$ nm device lie, as expected from section \ref{subsectionSection}, between the simulations for the $W=14$ nm device and for the planar $(001)$ FDSOI device. The larger discrepancies at $V_{\rm gt}\sim 0.2$ V might result from the lower accuracy of the extrapolations down to $L=0$ (enhanced device-to-device variability near the threshold and larger slope of the $R(L)$ data), and/or from missing Coulomb correlations in the simulations.}

The experimental data therefore support the main conclusions of this work about the dependence of the contact resistance on the gate bias and on the cross section. Such a $V_{\rm gt}^{-\beta}$ behavior, if not properly accounted for in compact models for the transistors, can strengthen the dependence of the mobility on gate length in short channels.\cite{Zilli07,Barral09,Rideau14}

\section{Conclusions}

We have computed the contact resistance $R_{\rm c}$ of trigate and FinFET devices using Non-Equilibrium Green's Functions. At low drain bias, $R_{\rm c}$ can represent a very large fraction of the total resistance of these devices, so that the design of the source/drain is as or even more important than the design of the channel in sub-20 nm technologies. The spacers between the heavily doped source/drain and the gate are the most resistive parts of the devices and dopant fluctuations under these spacers a major source of variability. The conductance under the spacers is typically limited by the poor electrostatic control over the charge density in these areas. The carrier density can indeed show a dip there, which enhances both ballistic and diffusive components of $R_{\rm c}$. As a consequence, the resistance of the spacers has a near $\propto 1/V_{\rm gt}$ dependence, which, if not properly accounted for, can partly explain the apparent dependence of the channel mobility on length in short devices.\cite{Zilli07,Barral09,Rideau14} We have investigated the impact of the channel width $W$ and height $H$, and the impact of the source/drain doping profiles on the device performances. The contact resistance $R_{\rm c}$ is more dependent on the cross sectional area $S=WH$ of the spacers than on the effective width $W_{\rm eff}=W+2H$ of the channel. Hence, devices such as FinFETs that maximize the effective width for a given cross sectional area tend to show larger normalized contact resistances $\bar{R}_{\rm c}=R_{\rm c}W_{\rm eff}$. Also, the contact resistance and variability dramatically increase when the cross sectional area of the channel is $\lesssim 50$ nm$^2$.  Finally, we have shown that defects at the interface between silicon and the spacer material act as surface dopants, which, depending on their charge, improve or degrade the contact resistance. 

This work was supported by the French National Research Agency (ANR project NOODLES). The NEGF calculations were run on the TGCC/Curie machine using allocations from PRACE and GENCI.


\begin{thebibliography}{71}
\bibitem [{Not()}]{NoteGeo}%
  \BibitemOpen
  \href@noop {} {}\bibinfo {note} {The system is therefore modeled as a
  periodic array of nanowires connected to thick source and drain films. The
  convergence with respect to $t_{\rm SD}$, $W_{\rm SD}$ and $L_{\rm SD}$ has
  been carefully checked.}\BibitemShut {Stop}%
\bibitem [{not()}]{noteDoping}%
  \BibitemOpen
  \href@noop {} {}\bibinfo {note} {The probability to find a dopant at at given
  point of the source is proportional to the background dopant concentration
  $N_{\rm back}$ shown in Fig. \ref{figDoping}. A distribution of point charges
  $N_{\rm point}$ is picked randomly, and mixed with $N_{\rm back}$ near the
  edge $z=z_{\rm min}$ of the simulation box: $N=\alpha(z) N_{\rm
  back}+[1-\alpha(z)] N_{\rm point}$, where $\alpha(z)^{-1}=1+10^{(z-z_{\rm
  mix})/\lambda_{\rm mix}}$, with $z_{\rm mix}=z_{\rm min}+4$ nm and
  $\lambda_{\rm mix}=0.66$ nm.}\BibitemShut {Stop}%
\bibitem [{Not({\natexlab{a}})}]{NoteBall}%
  \BibitemOpen
  \href@noop {} {} ({\natexlab{a}}),\ \bibinfo {note} {$\bar{R}_0^{\rm NW}$ can
  either be obtained from a ballistic NEGF calculation in a homogeneous
  nanowire or from the band structure of that nanowire (see, e.g., Refs.
  \onlinecite{Natori94} and \onlinecite{Rahman03}).}\BibitemShut {Stop}%
\bibitem [{Not({\natexlab{b}})}]{NoteVt}%
  \BibitemOpen
  \href@noop {} {} ({\natexlab{b}}),\ \bibinfo {note} {$\hat{V}_{\rm t}$ can be
  slightly different from $V_{\rm t}$ defined in paragraph
  \ref{subsectionMethodology}, as the former is based on a density criterion,
  and the latter on a current criterion.}\BibitemShut {Stop}%
\bibitem [{not({\natexlab{a}})}]{noteplanar}%
  \BibitemOpen
  \href@noop {} {} ({\natexlab{a}}),\ \bibinfo {note} {the planar $(001)$ FDSOI
  device used for comparison in Fig. \ref{figRcsection} is a 8 nm thick Si film
  on a 25 nm thick BOX with the same gate stack and surface roughness
  parameters as the trigate devices (namely $\Lambda=1.5$ nm and $\Delta=0.25$
  nm on the bottom interface with the BOX, $\Lambda=1.5$ nm and $\Delta=0.35$
  nm on the top interface with the gate stack). The planar $(110)$ FDSOI device
  is the same, but with $\Lambda=2.0$ nm and $\Delta=0.45$ nm on both top and
  bottom interfaces. The planar $(110)$ double-gate device is, likewise, a
  symmetric stucture with a 8 nm thick film, and $\Lambda=2.0$ nm,
  $\Delta=0.45$ nm on both interfaces.}\BibitemShut {Stop}%
\bibitem [{not({\natexlab{b}})}]{noteContact}%
  \BibitemOpen
  \href@noop {} {} ({\natexlab{b}}),\ \bibinfo {note} {the source and drain can
  be modeled as $90\times 90$ nm pads with contact resistivities $\rho_{\rm
  c}\simeq 2\times 10^{-8}$ $\Omega$.cm$^2$. For a typical effective width
  $W_{\rm eff}=50$ nm, these pads therefore make a contribution to the contact
  resistance ${\bar R}_{\rm c}^\prime\simeq 24$ $\Omega.\mu$m.}\BibitemShut
  {Stop}%
\end{thebibliography}
\providecommand{\noopsort}[1]{}\providecommand{\singleletter}[1]{#1}%

\end{document}